# Nanodiamond-based spatial-temporal deformation sensing for cell mechanics

Yue Cui[2,1,+], Weng-Hang Leong[3,1,+], Guoli Zhu[1,+], Ren-Bao Liu[1,4,5,6,*], Quan Li[1,4,5*]

[+]These authors contributed equally: Yue Cui, Weng-Hang Leong, Guoli Zhu

[*]Correspondence and requests for materials should be addressed to R.-B.L. (email: rbliu@cuhk.edu.hk) or to Q.L. (email: liquan@cuhk.edu.hk)

1. Department of Physics, The Chinese University of Hong Kong, Shatin, New Territories, Hong Kong, China

2. Quantum Science Center of Guangdong-Hong Kong-Macao Greater Bay Area, Shenzhen, China

3. Department of Engineering Science, Faculty of Innovation Engineering, Macau University of Science and Technology, Taipa, Macao 999078, China

4. Centre for Quantum Coherence, The Chinese University of Hong Kong, Shatin, New Territories, Hong Kong, China

5. The Hong Kong Institute of Quantum Information Science and Technology, The Chinese University of Hong Kong, Shatin, New Territories, Hong Kong, China

6. New Cornerstone Science Laboratory, The Chinese University of Hong Kong, Shatin, New Territories, Hong Kong, China

**ABSTRACT**

Precise assessment of the mechanical properties of soft biological systems at the nanoscale is crucial for understanding physiology, pathology, and developing relevant drugs. Conventional atomic force microscopy (AFM)-based indentation methods suffer from uncertainties in local tip-sample interactions and model choice. This can be overcome by adopting spatially resolved nonlocal deformation sensing for mechanical analysis. However, the technique is currently limited to lifeless/static systems, due to the inadequate spatial or temporal resolution, or difficulties in differentiating the indentation-induced deformation from that associated with live activities and other external perturbations. Here, we develop an innovative dynamic nonlocal deformation sensing approach allowing both spatially and temporally resolved mechanical analysis, which achieves a tens of microsecond time-lag precision, a nanometer vertical deformation precision, and a sub-hundred nanometer lateral spatial resolution. Using oscillatory nanoindentation and spectroscopic analysis, the method can separate the indentation-caused signal from random noise, enabling live cell measurement. Using this method, we discover a distance-dependent phase of surface deformation during indentation, leading to the disclosure of surface tension effects (capillarity) in the mechanical response of live cells upon AFM indentation. A viscoelastic model with surface tension is used to enable simultaneous quantification of the viscoelasticity and capillarity of cell. We show that neglecting surface tension, as in conventional AFM methods, would underestimate the liquid-like characteristics and overestimate the apparent viscoelastic modulus of cells. The study lays down a foundation for understanding a broad range of elastocapillarity-related interfacial mechanics and mechanobiological processes in live cells.

# I. INTRODUCTION

Cell mechanical properties are tightly associated with a variety of cellular processes, including division [1], differentiation [2], diffusion [3], and motility [4], and are important emerging biomarkers for cancer detection, diagnosis, and classification [5–7]. Various methods such as atomic force microscopy (AFM) and magnetic/optical tweezer, which allows dynamic measurement of mechanical forces and deformations at the single cell level, have advanced our understanding of the frequency-dependent cell viscoelasticity and rheological behaviors over a wide frequency range [5,7,8]. However, these techniques based on force-deformation measurement at the local indentation position still face several limitations in actual cell mechanical characterization. On one hand, relating the data to quantitative mechanical properties relies on correct modeling, and the complications of local contact make the data interpretation often ambiguous [9,10]. On the other hand, the methods lack the information of the nonlocal (spatial) response of cell to the local loading. The nonlocal information could be useful for observing important mechanical information related to cell interface such as the capillary action and surface tension effect [11,12].

Nonlocal deformation measurement, in which the deformation is measured at a location away from where the deformation is induced (e.g., by nano-indentation), is advantageous for analyzing the intrinsic mechanical properties of materials, since it is unaffected by complications at the local contacts [13]. The last decade has seen revolutionary advances in development of methods for measuring deformation with nonlocal information, based on optical imaging [14,15] or deformation reconstruction from sensor rotation [10,11]. Recent works in this emerging field have shown that the usually overlooked surface tension in solid can play a central role in the mechanics of soft solids such as polymer gels [12] and elastomers [16], and can affect the evaluation of the elastic modulus of fixed cells by AFM indentation [11]. Despite the importance of interfacial mechanics in regulating cellular processes [17–19], the influence of surface tension (capillarity) in the mechanical responses of live cells upon AFM indentation remains unclear, due to the challenges in applying the nonlocal deformation sensing methods in live cells. The optical imaging method does not have sufficient precision and resolution, where nanometer precision for deformation measurement is needed for less-invasive (hence shallow) indentation on the live systems and nanoscale lateral spatial resolution is required since the critical elastocapillary lengths related to the elastocapillary phenomena of biological systems, defined as the ratio of surface tension to bulk Young's modulus, are usually in the order of micron [11]. The nonlocal deformation sensing methods based on nanodiamond (ND) rotation sensing features nanometer precision as well as sub-hundred nanometer spatial resolution [10], but the technique is currently restricted to lifeless and static systems, since it cannot differentiate indentation-induced material deformation from that originating from live-associated activities [11]. It also lacks the information about the temporal mechanical response of materials and is incapable of capturing material viscoelasticity.

In this work, we develop a diamond-based dynamic nonlocal deformation sensing scheme with high precision, high spatial and temporal resolution, which enables both spatially and temporally resolved mechanical analysis of soft materials and live cells. We overcome the limitation of the diamond-based deformation sensing to static systems by combining oscillatory AFM nanoindentation [5,8] with frequency-domain measurement of the induced deformation. The dynamic deformation is converted to periodically variations in the orientation of the nitrogen-vacancy (NV) centers in diamond through the diamond sensors pre-anchored on the material surface. Based on the established quantum sensing protocols for orientation measurement of NV centers, we develop a fast rotation sensing method using a two-point optically detected magnetic resonance (ODMR) measurement with high duty ratio. We realize phase-sensitive detection of the ND rotation by synchronizing the AFM indentation and fast rotation measurement. The signals at the oscillation frequency single out the indentation-induced deformation from that induced by other sources such as rotation diffusion and cell activities. An additional benefit of the AC measurement is that it filters out the background fluorescence noise, and therefore enhances the signal-to-noise ratio and enables the deformation measurement in complex live systems.

We demonstrate the scheme by measuring viscoelastic polydimethylsiloxane (PDMS) films and live MCF-7 cells. Through the phase-sensitive spectral analysis, we achieve $0.01\,\pi$ precision in phase lag measurement (tens of microsecond time-lag precision), ~2 nm vertical deformation precision, and sub-hundred nanometer lateral spatial resolution in the dynamic nonlocal deformation mapping. Our experiments lead to the first discover of a distance-dependent phase lag of the nonlocal deformation during the oscillatory indentation. The phase lag reveals the elastocapillary effects in the mechanical response, indicating that the bulk viscoelasticity (dominant at larger scales) and the surface tension (dominant at smaller scales) are competing factors when the lateral length scales of the deformation are comparable to the elastocapillary length of the material (typically in the order of microns for soft materials). The interplay between the bulk and surface responses in the nonlocal deformation enables simultaneous assessment of the surface tension (capillarity) and the viscoelasticity (quantified by a frequency-dependent complex modulus $\tilde{E}^*(f) = \tilde{E}'(f) + i\tilde{E}''(f) = |\tilde{E}^*(f)|e^{i\delta_{\text{loss}}}$, where the real part $\tilde{E}'(f)$ and the imaginary part $\tilde{E}''(f)$ are the storage and loss moduli, respectively, and the loss angle $\delta_{\text{loss}}$ varies between 0 and $\pi/2$ with 0 representing a complete solid and $\pi/2$ a complete liquid). We show that an overlook of the surface tension, as in conventional local AFM-indentation measurement using classical contact models, would lead to an overestimation of the magnitude ($|\tilde{E}^*(f)|$) and an underestimation of the liquid-like characteristics ($\delta_{\text{loss}}$) of cell viscoelasticity (since in local measurement the surface tension adds effectively to the ratio of the real part to the imaginary part of the complex modulus). Our result constitutes the first unambiguous measurement of the elastocapillary effect in AFM indentation of live cells, underscoring the crucial role of surface tension in the mechanical response of cell. The nanodiamond quantum sensor is also demonstrated as a useful tool for quantifying the intrinsic mechanical properties of soft, complex biological systems.

**II. RESULTS**

**A. Dynamic nonlocal deformation measurement based on ND rotation sensing**

Figure 1 illustrates the scheme for dynamic nonlocal deformation measurement. We used a home-built AFM-confocal correlated microscopy setup (for details, see Methods and Supplementary Figure S1). We applied shallow indentations (500-800 nm indentation depth) to soft materials [see Fig. 1(a)]. Upon a certain holding force, we added a modulation [Fig. 1(b)] to induce oscillatory indentation at a fixed frequency [Fig. 1(c)]. The oscillatory indentation, through surface deformations, caused oscillatory rotations of the NDs pre-attached on the sample surfaces [Fig. 1(d)]. Through phase-sensitive spectral analysis of the ND rotations (see Supplementary Figure S2 for the measurement sequence), we isolated the indentation-induced deformation signals, from those originating from fluctuating environmental parameters and/or cellular activities (in the case of live cells).

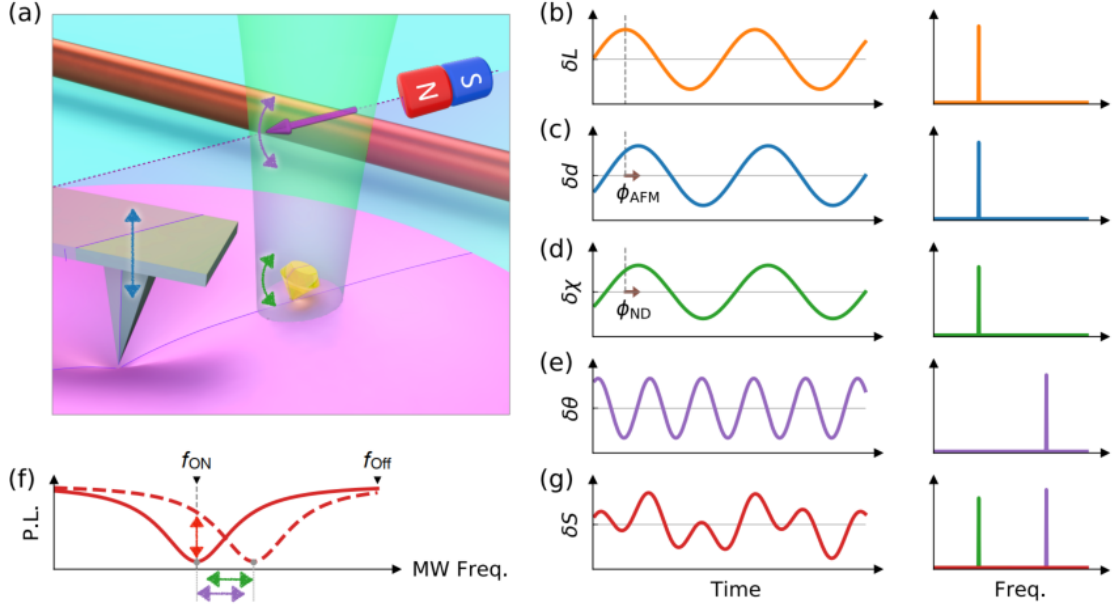

FIG. 1. Dynamic nonlocal deformation reconstruction using nanodiamond (ND) rotation sensing. (a) Schematic of an atomic force microscope (AFM) tip imposing an oscillatory indentation on a soft material. NDs are attached on the surface for dynamic rotation sensing by optically detected magnetic resonance (ODMR). (b) to (e), Time evolution (left) and the corresponding spectra (right) of the loading $L$, indentation depth $d$, ND rotation angle $\chi$ and magnetic field rotation angle $\theta$. The brown arrows in (c) and (d) indicate the local and nonlocal phase lags ($\phi_{\text{AFM}}$ and $\phi_{\text{ND}}$). (f) Schematic of the two-point ODMR method. The red lines show the changes of ODMR spectra with consequent oscillations in the emission of the ND (indicated by red arrow) due to the rotation of ND (indicated by green arrow) and that of magnetic field (indicated by purple arrow). The on- and off-resonance frequencies ($f_{\text{on}}$ and $f_{\text{off}}$) of the nitrogen-vacancy (NV) centers spin transitions are indicated by the black triangles. (g) Time evolution (left) and the corresponding spectrum (right) of the two-point ODMR signal $S$. The corresponding peaks in the spectrum induced by the oscillatory loading and the rotating magnetic field are colored by green and purple, respectively.

For the phase-sensitive detection of dynamic ND rotation, two required techniques are developed here. First, an accurate synchronization between the force loading and the nonlocal deformation measurement was achieved by using a trigger signal to align the start of data recording (refer to Methods and Supplementary Figure S3 for details), which is important for precise measurement of the phase lag between force and deformation. Second, we developed a fast rotation sensing scheme based on a two-point ODMR method with high duty ratio, to overcome the limit of time resolution (in the order of seconds) of the previous ND rotation measurement using continuous-wave ODMR [20], which is inadequate for tracking the dynamic nonlocal deformation of the sample. The method is enabled by applying an external magnetic field with known oscillatory rotation. This magnetic field serves as a reference for calibrating the rotation of the ND, eliminating the need to record the full ODMR spectra to obtain a complete set of rotation information [10]. Such calibration is realized by the dependence of the ground state spin resonance frequency of NV centers ($f_\pm$) on the magnetic field components along the NV axis with $f_\pm \approx D \pm \gamma_e \mathbf{B}(t) \cdot \mathbf{n}_{NV}(t)$, where $D$ is the zero-field splitting, $\gamma_e$ is the electron gyromagnetic ratio [21,22]. Changing the orientation of ND [represented by $\mathbf{n}_{NV}(t)$] and that of the magnetic field [represented by $\mathbf{B}(t)$] have equivalent effects on the resonance frequency $f_\pm$. Assuming a known rotation axis of the ND upon indentation [10,11], we applied a known rotating magnetic field with the same rotation axis but a different oscillating frequency [refer to Figs. 1(d) and 1(e)]. This allows us to calibrate the magnitude of the ND rotation caused by indentation (for details of the magnetic field control, see Methods and Supplementary Figure S4).

In the two-point ODMR method, as depicted in Fig. 1(f), the photon emission is collected at an on-resonance microwave (MW) frequency and an off-resonance frequency to normalize fluorescence fluctuation (refer to Methods and Supplementary Figure S2). The coaxial oscillatory rotation of the ND and that of the magnetic field induce peak shift of ODMR and consequent oscillations in the emission of the ND. Figure 1(g) (left) illustrates the ODMR signal of the ND as a function of time, displaying a beat signal resulting from the two frequency components. The spectrum obtained by Fourier transformation in Fig. 1(g) (right) exhibits two distinct peaks corresponding to the rotation of the ND and that of the magnetic field. Since the rotating magnetic field is pre-calibrated and known (refer to Supplementary Figure S4), both the amplitude and phase of the ND rotation can be determined from these peaks.

To further enhance the signal-to-noise ratio of the measurement, we utilize not only one resonance peak in ODMR but simultaneously incorporate the four resonance peaks originating from the four NV orientation classes in the ND, by applying a mixture of the four on-resonance MWs in the ODMR detection process (see Supplementary Figure S5 for details). This approach enhances the contrast of the two-point ODMR measurement by a factor of approximately four.

**B. Proof-of-the-concept demonstration: the spatial-temporal mechanical response of PDMS**

As a proof-of-concept demonstration, we conducted measurements on the dynamic nonlocal deformation of a PDMS film (see Methods for sample preparation). Figure 2(a) presents the AFM topography image of a PDMS film with NDs anchored on its surface. We performed sequential AFM indentations at 20 different locations near an ND (see Methods and Supplementary Figure S6 for more information). The locations were divided into two groups on two sides of the ND sensor, and the displacements from the AFM indentation spots to the ND were set parallel to the projection of the external magnetic field on the *xy*-plane (for more information, refer to Supplementary Figure S4). This alignment ensured that the magnetic field and ND shared the same rotation axis as the deformation is axisymmetric under the loading by an axisymmetric tip [10,11].

Prior to the indentations, we recorded a ODMR spectrum to determine the resonance frequencies, as shown in Fig. 2(b). For each measurement, synchronized AFM oscillatory indentation and two-point ODMR collection were performed. Figure 2(c) illustrates one set of AFM and ODMR data as a function of time. Detailed information on the corrections for laser reflection and hydrodynamic drag on the AFM cantilever can be found in the Methods section, along with illustrations in Supplementary Note 1 and Figures S7 and S8. The corresponding data in the frequency domain, obtained through Fourier transformation (FT), are presented in Fig. 2(d). The top two panels of Figs. 2(c) and 2(d) display the local indentation data, including loading and indentation depth, obtained from AFM measurements. The bottom panel depicts the two-point ODMR signal of the ND, revealing the beat of two frequency components at 10 Hz and 25 Hz, corresponding to the loading and magnetic field modulation, respectively. The amplitude and phase of each oscillation were derived from the FT signal at the corresponding frequency (see Methods for details on data analysis). By comparing the phases, we obtained the phase lag of the local deformation $\phi_{\text{AFM}}$ and nonlocal deformation $\phi_{\text{ND}}$ relative to the loading.

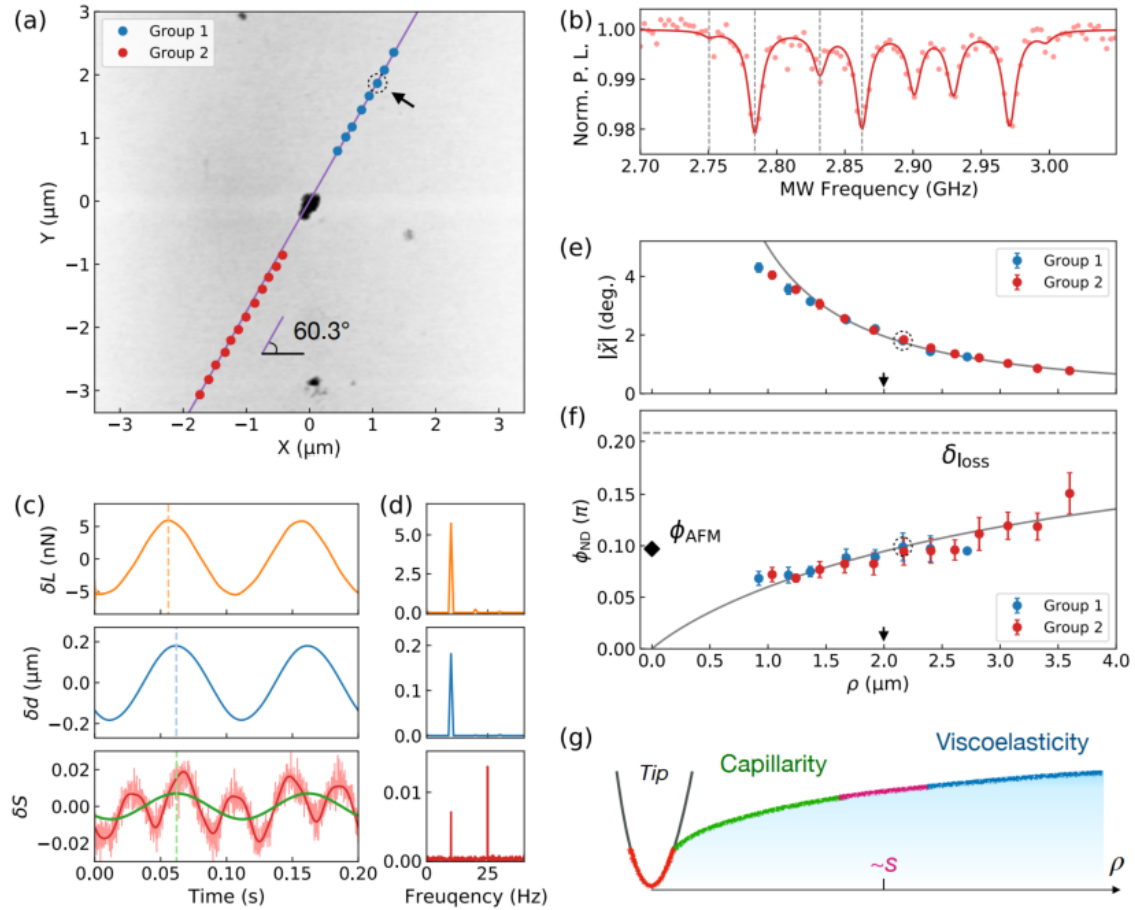

FIG. 2. Dynamic nonlocal deformation measurement using ND rotation sensing on polydimethylsiloxane (PDMS). (a) AFM image of a typical PDMS surface with an ND located at the center (the origin). The red

and blue dots represent the indentation locations of the AFM tip on the upper right (Group 1) and the lower left (Group 2) of the ND. The purple line indicates the direction of the external magnetic field B (60.3° from the x-axis). (b) The ODMR spectra of the ND under the magnetic field obtained before the indentations. The four resonance frequencies are indicated by the grey dashed lines. (c) The time-dependent data and (d) the Fourier transform (FT) of the loading $L$, the depth $d$ of the AFM tip and the two-point ODMR signal $S$ at the indentation location indicated by the black arrow in (a). The red solid curve is the simulation of the beat signal based on the deduced amplitude and phase, while the green line is the extracted 10 Hz signal of ND rotation induced by indentation. The dashed colored lines in (c) show the phase shifts of the local depth and the extracted nonlocal ND rotation relative to the loading. (e) The amplitude ($|\tilde{\chi}|$) and (f) the nonlocal phase lag ($\phi_{ND}$) of the oscillatory rotation angle as functions of the distance $\rho$ between the indentation location and the ND. The data points are averages over ten repeated measurements executed at each location. The fitting results of the linear viscoelastic model with surface tension effect are plotted by the grey lines, where the elastocapillary length $|\tilde{s}|$ is pointed by the black arrows in (e) and (f). The local phase lag $\phi_{AFM}$ deduces from the AFM data is drawn by the rhombus at $\rho = 0$, while the bulk loss angle $\delta_{loss}$ is indicated by the grey dashed line. The error bars in (e) and (f) are the standard derivation of the repeated measurements. (g) Schematic of the elastocapillary effect in AFM indentation.

Figures 2(e) and 2(f) display the oscillation amplitude $|\tilde{\chi}|$ of the ND rotation angle and the phase lag $\phi_{ND}$ between loading and nonlocal deformation at various distances $\rho$ between the indentation positions and the ND. The two sets of data, obtained from the two sides of the ND [see Fig. 2(a)], exhibit good agreement, underscoring the consistency of the measurements on the homogeneous sample. The rotation amplitude decreases with increasing distance $\rho$, while the phase lag increases with distance. This spatially dependent phase lag indicates the interplay between capillarity (characterized by the time-independent surface tension $\tau_0$ at the material-liquid interface) and viscoelasticity (characterized by the frequency-dependent complex modulus $\tilde{E}^*(f)$ of the material). This interplay becomes significant when measured at a lateral length scale comparable to the elastocapillary length $\tilde{s}(f) \equiv 2\tau_0/\tilde{E}^*(f)$, as depicted in Fig.

2(g) [12]. In the regime where $\rho \gg |\tilde{s}|$, the time-dependent bulk viscoelasticity dominates, with a phase lag determined by the material's loss angle $\delta_{\text{loss}}$. Conversely, in the deformation regime where $\rho \ll |\tilde{s}|$, the time-independent capillary force dominates, resulting in no phase lag.

According to Saint-Venant's principle[1], the nonlocal deformation far away from the indentation point (with distance $\rho > 800$ nm, much larger than the tip radius of approximately 25 nm) is independent of the local factors of the AFM indentation and can be well approximated by the deformation upon point loading [23]. Therefore, a linear viscoelastic model based on a point loading accounting for the surface tension effect [23,24] is used to explain the observed nonlocal mechanical response of the PDMS (see Methods and Supplementary Figure S9 for details of the model). The viscoelastic properties and surface tension of the PDMS, measured at 10 Hz, are deduced to be $|\tilde{E}^*| = 7.6(4)$ kPa, $\delta_{\text{loss}} = 0.21(2)\,\pi$, and $\tau_0 = 7.4(6)$ mN m$^{-1}$, which are consistent with previous reports [16,25]. The results indicate an elastocapillary length of $|\tilde{s}| = 2$ μm, as indicated by the black arrows in Figs. 2(e) and 2(f). For comparison, the derived $\delta_{\text{loss}}$ of the PDMS is plotted in Fig. 2(f) with a gray dashed line, and the local phase $\phi_{\text{AFM}}$ obtained from conventional AFM dynamic measurements is represented by the black diamond at $\rho = 0$. The local phase at $\rho = 0$ deviates from the fitting results of zero phase lag (when only the static surface tension is considered). This deviation is induced by the local factors of the AFM indentation, where the point-loading approximation is invalid.

To further demonstrate the method's capability in evaluating the frequency-dependent viscoelasticity, the nonlocal deformation measurements were repeated using different loading modulation frequencies ranging from 10 to 80 Hz (for detailed information, refer to Supplementary Figures S6, S10 and S11). In Figs. 3(a) and 3(b), the amplitude of ND rotation, normalized by the constant $\chi_L = |\tilde{L}(\omega)|/[2\pi\tau_0|\tilde{s}(\omega)|]$, and the nonlocal phase, normalized by $\delta_{\text{loss}}$, are plotted as functions of the rescaled distance $\rho/|\tilde{s}(\omega)|$.

The parameters $\chi_L$ and $\delta_{\text{loss}}$ are obtained by fitting the nonlocal deformation data using the point-loading model with elastocapillary effects, and the fitting results are represented by the gray lines. The rescaled results obtained at different frequencies approximately align along the same line, following the approximately universal function provided by the viscoelastic model with capillary effects (see Methods and Supplementary Figure S9). The standard deviation of the oscillation amplitude $|\tilde{\chi}|$ was $\sigma_{|\tilde{\chi}|} = 0.46°$, which defines a nanometer precision of deformation reconstruction in the loading direction ($\sigma_{|\tilde{z}|} = 1.7$ nm) (see Supplementary Figure S12). Meanwhile, the standard deviation of the phase lag $\phi_{\text{ND}}$ was $\sigma_{|\phi_{\text{ND}}|} = 0.01\,\pi$, which defines a tens of microsecond precision of time-lag measurement (63 us for the 80 Hz oscillation).

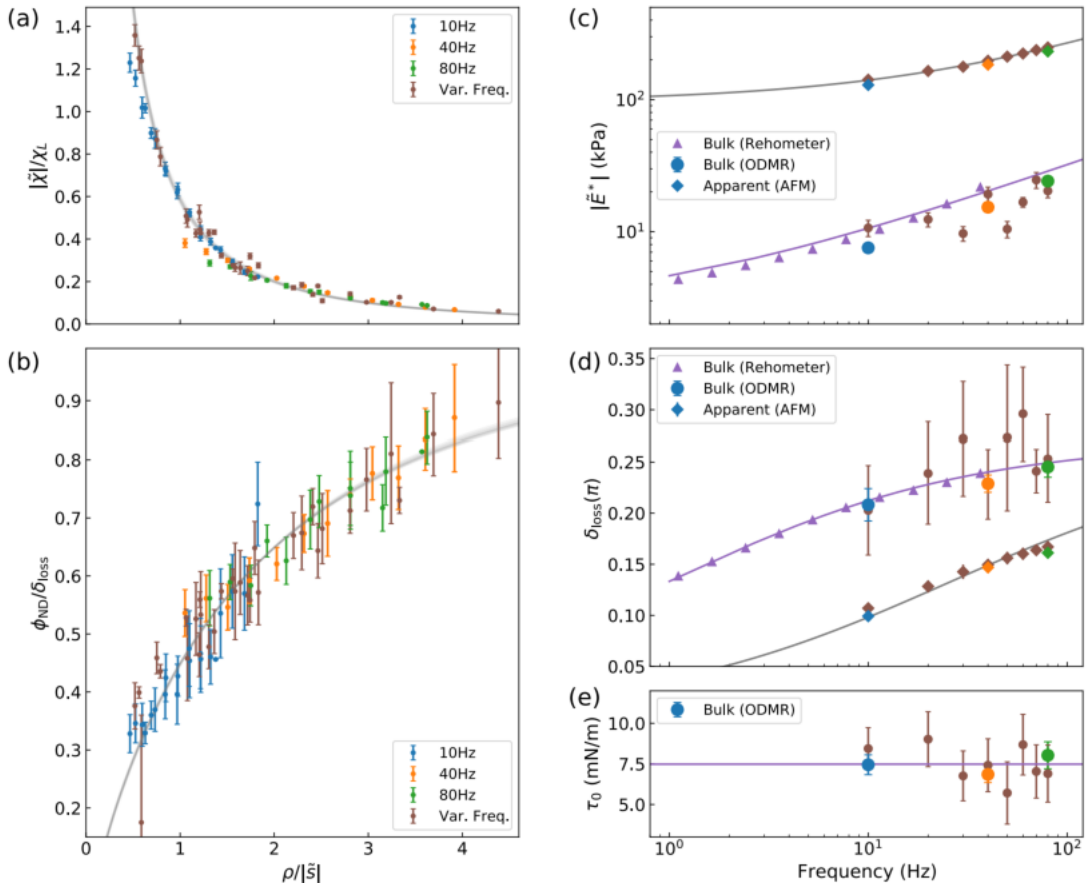

FIG. 1. Evaluation of the complex modulus and surface tension of the PDMS film. (a) The rescaled

amplitude of the oscillatory rotation angle ($|\tilde{\chi}|/\chi_L$) and (b) the normalized nonlocal phase lag ($\phi_{ND}/\delta_{loss}$) of the oscillatory rotation angle as functions of the rescaled distance $\rho/|\tilde{s}|$ between the indentation spot and the ND. The simulation results of the linear viscoelastic model with surface tension effect are plotted by the grey lines. The error bars in (a) and (b) are similar to those of Figs. 2(e) and (f). (c) The evaluated magnitude $|\tilde{E}^*|$ and (d) the loss angle $\delta_{loss}$ of the complex modulus as functions of frequency. The purple triangles (colored rhombuses) plot the magnitude and loss angle of the bulk (apparent) complex modulus of PDMS measured by rheometer (AFM). The lines are the fitting results of the bulk and apparent elastic moduli by using the power-law function as illustrated in the main text. (e) The surface tension $\tau_0$ deduced in the different experiments. The line indicates the mean value. The error bars in (c) to (e) are fitting errors.

The mechanical properties of the PDMS at different frequencies were deduced by fitting the nonlocal deformation data corresponding to the respective frequencies. The deduced complex modulus ($|\tilde{E}^*|$, $\delta_{loss}$) and surface tension ($\tau_0$) are shown as functions of the loading modulation frequency, represented by colored circles in Fig. 3(c) to 3(e). Both the modulus $|\tilde{E}^*|$ and the loss angle $\delta_{loss}$ increase with frequency, while the surface tension exhibits no apparent frequency dependence with a mean value of 7.5 mN m$^{-1}$, which is in agreement with a recent report [16]. The bulk viscoelastic properties of the PDMS were also measured with a rheometer (see Methods), represented by the purple triangles in Fig. 3(c) and 3(d). As PDMS exhibits scale invariance as cross-linked polymer networks, the bulk complex moduli were fitted using a power-law function $\tilde{E}^*(f) = E_0^*(1 + (2\pi i \tau f)^n)$, with the static modulus $E_0^* = 2$ kPa, the characteristic time $\tau = 0.08$ s, and the exponent n = 0.55. The mechanical properties evaluated from the dynamic nonlocal deformation measurements are consistent with the results from the rheometer and also with the literature [25].

For comparison, the apparent complex modulus $\tilde{E}_a^*$ was calculated using the local dynamic loading and deformation data (refer to Supplementary Figure S13) by the conventional Hertz-Sneddon model [26,27], with the surface tension effect neglected (see Methods). The apparent loss angle $\delta_{apparent}$ by conventional

AFM method was obtained from the measured local phase lag of the indentation depth relative to the loading, i.e., $\phi_{\text{AFM}}$. The apparent modulus $\tilde{E}_a^*$ and the loss angle $\delta_{\text{apparent}}$ are represented by solid diamonds in Fig. 3(c) and 3(d), respectively, which deviate significantly from the bulk viscoelastic properties obtained through nonlocal measurements and oscillatory rheology (i.e., $|\tilde{E}_a^*| > |\tilde{E}^*|$ and $\delta_{\text{apparent}} < \delta_{\text{loss}}$). These deviations suggest that neglecting the surface tension in conventional methods would lead to an overestimation of the apparent modulus [Fig. 3(c)] and an underestimation of the loss angle [Fig. 3(d)]. The same measurements were repeated on additional PDMS samples, yielding similar extracted mechanical properties, as shown in Supplementary Figure S14.

**C. Spatial-temporal mechanical response of live cells**

Live cells are complex active materials exhibiting both solid-like elastic and liquid-like viscous features. Our method was further tested on living systems by measuring the spatial-temporal mechanical response of live MCF-7 cells. We measured the rotation of a single ND anchored to the plasma membrane upon AFM indentation (see Methods for cell culture, cell status, and sample preparation). Figures 4(a) and 4(b) present side and top views, respectively, of confocal fluorescence images of the MCF-7 cell, with a white arrow indicating the presence of an ND on the cell membrane. As the orientation of the ND on the cell membrane changed over time due to cellular activities [20], we collected the ODMR spectrum before each indentation to obtain the four transition frequencies, and then performed the two-point ODMR measurements (refer to Supplementary Figure S16 and S17 for an example of the data).

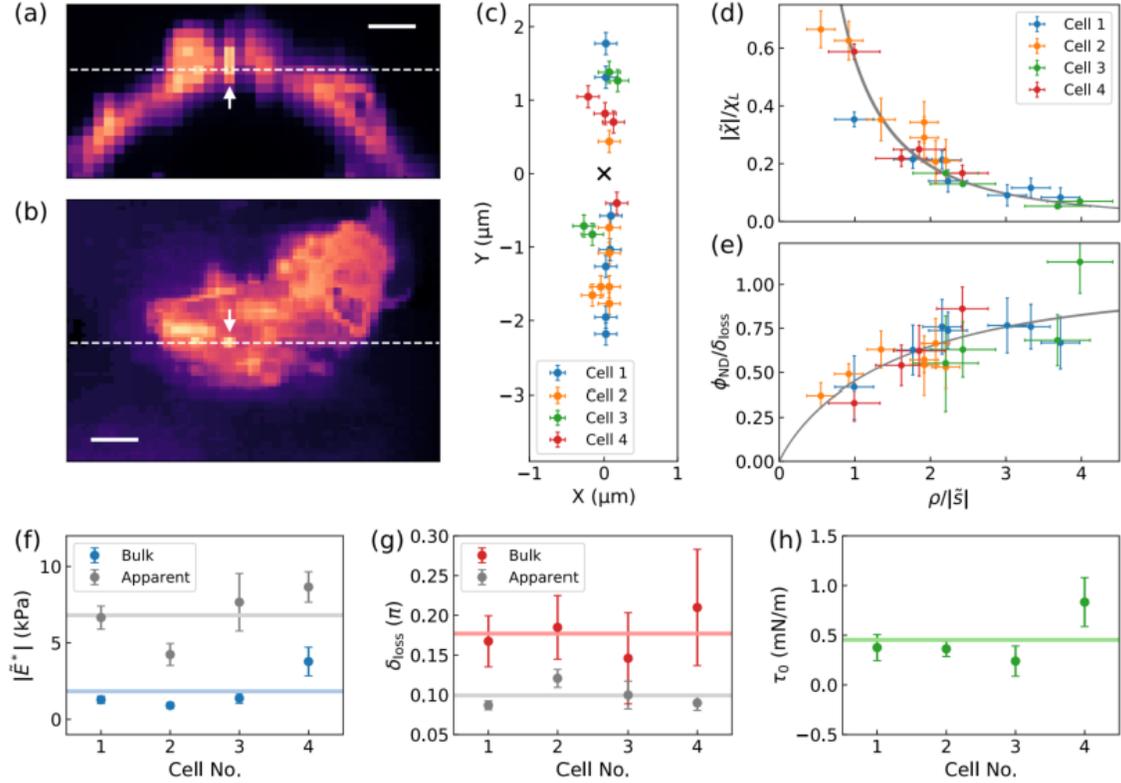

FIG. 4. Elastocapillary effect on live cells. (a) The *xz*-cross section and (b) The *xy*-cross section of the confocal fluorescence images of the MCF-7 cell with an ND on the cell membrane as indicated by the white arrows. The white dashed lines show the cutting positions of the cross sections. The scale bars are 5 μm. (c) The displacements of the indentation spots from the NDs (denoted by the black cross at the origin). The AFM indentations were performed along the *y*-axis, parallel to the direction of the external magnetic field. (d) The rescaled amplitude ($|\tilde{\chi}|/\chi_L$) and (e) the normalized phase lag ($\phi_{\text{ND}}/\delta_{\text{loss}}$) as functions of the rescaled distance $\rho/|\tilde{s}|$. The simulation results are plotted by the grey lines. The error bars in (c) and *x*-axis error bars in (d) and (e) are the optical resolution ($\pm 150$ nm). The *y*-axis error bars in (d) and (e) are estimated form the signal-to-noise ratio of the FT data of the two-point ODMR signal. (f) The evaluated magnitude $|\tilde{E}^*|$ and (g) the loss angle $\delta_{\text{loss}}$ of the bulk (colored dots) and apparent (grey dots) complex modulus of the four live cells. (h) The surface tension $\tau_0$ of the four cells. The lines in (f) to (h) indicate the mean value. The error bars in (f) to (h) are fitting errors.

The normalized amplitude and the nonlocal phase of ND rotation upon AFM indentation with modulation frequency of 40 Hz are plotted as functions of the rescaled distance in Figs. 4(d) and 4(e). The data are well fitted by a viscoelastic model that includes the surface tension effect, allowing estimation of the mechanical properties of the MCF-7 cell as $|\tilde{E}^*| = 1.0(2)$ kPa, $\delta_{\text{loss}} = 0.17(3)\,\pi$ and $\tau_0 = 0.4(1)$ mN m$^{-1}$. Similar measurements were conducted on three additional MCF-7 cells, and their nonlocal deformation data are also presented in Figs. 4(c) to 4(e). The normalized data from all cells were well-described by the viscoelastic model with the inclusion of surface tension. The observed spatially non-uniform phase of nonlocal deformation suggests that the surface tension significantly influences the mechanical response of live cells during AFM shallow indentation. Figs. 4(f), 4(g) and 4(h) display the deduced complex modulus ($|\tilde{E}^*|$ and $\delta_{\text{loss}}$) and surface tension ($\tau_0$) of the four live cells based on the ND rotation data. The deduced complex modulus is consistent with the previous reports measured by AFM [28,29], and the deduced surface tension agrees with the range of the membrane tension and cortex tension measured in the literature [30,31]. As a comparison, the apparent complex moduli ($\tilde{E}_a^*$) of the cells were evaluated from the local deformation data (see Supplementary Figure S18) using the Hertz-Sneddon model, represented by grey dots in Figs. 4(f) and 4(g). Considering surface tension in the contact model is also found to be crucial for accurately assessing cell mechanical properties, similar to the previously discussed results obtained from PDMS. Neglecting surface tension, as done in conventional AFM indentation methods, would lead to an overestimation of the apparent modulus of the cell and an underestimation of the loss angle, indicating an underestimation of the liquid-like characteristics of the cell.

**III. CONCLUSIONS**

In summary, we have developed a dynamic nonlocal deformation sensing scheme for spatial-temporal mechanical analysis using modulated AFM nanoindentation and frequency-specific ND rotation sensing. We successfully demonstrated frequency-dependent dynamic nonlocal deformation mapping with high precision, high spatial and temporal resolution on soft viscoelastic PDMS and live cells. The measurements on cells demonstrate the robustness of the method against background deformation noise caused by cellular activities. Our method leads to the first measurement of the spatially dependent phase lagging of nonlocal deformation, resulting from the interplay between viscoelasticity and capillarity. Importantly, our nonlocal measurement approach allows for unambiguous quantitative analysis of the intrinsic mechanical properties of soft materials, including the viscoelastic complex modulus and surface tension. Our results highlighted the elastocapillary effect at the material/liquid interface of viscoelastic materials, such as polymers and live cells, during AFM indentation. We showed that neglecting the surface tension effect in contact mechanics leads to an overestimation of the apparent complex modulus and an underestimation of the loss angle of the materials.

The important insights into the competition between viscoelasticity and capillarity bring new inspiration to studies of cell mechanics. While the surface tension of live cells [31] (with a typical value of $10^{-1}$ mN m$^{-1}$) does not change with frequency, the elastic modulus of cells increases from $10^{-2}$ to $10^1$ kPa in the frequency range of $10^{-1} - 10^5$ Hz [5]. Consequently, the elastocapillary length of cells would change significantly from tens of microns to tens of nanometers. Since the deformation length scale is frequency-independent (typically $10^{-1} - 10^0$ μm), the large variation in the elastocapillary length at different frequencies could cause bias on the evaluation of the frequency-dependent rheological characteristics of cells, especially in the low frequency regime. By considering the surface tension effect, it would be interesting to revisit some previous discussions on cell mechanics based on the local measurement of apparent elastic modulus [5,32], for examples, the explanations of the weak power-law rheological behaviors of cell at low frequencies [33]. Overall, the nanodiamond-based dynamic nonlocal deformation sensing offers a unique tool and opens up new possibilities for studying the mechanics of live cells and other soft bio-relevant viscoelastic materials, with potential applications ranging from investigating cell rheological behaviors to establishing cancer diagnostic mechanical fingerprints.

## IV. METHODS

### A. AFM-confocal correlated microscopy

The measurements were carried out on a home-built confocal-AFM correlation microscope (see Supplementary Figure S1 and also Supplementary Note 1 of Ref. [ [10]] for more details). We used the laser scanning confocal system for imaging and ODMR measurements. The spins of NV centers were pumped by a 532 nm laser and manipulated by microwave delivered via a copper wire antenna (20 μm diameter). A BioScope Resolve AFM (Bruker) was used for surface imaging and for nanoindentation.

*Spatial correlation:* For the PDMS experiments, the spatial correlation between the AFM and the confocal microscope was established by overlapping the AFM image and the fluorescence image of NDs on the samples [10]. The AFM imaging provides the coordinates of different indentation spots with spatial resolution limited by the tip radius (∼25 nm, DNP-10-A, Bruker). For the live cells experiments, the position of the AFM tip in the confocal image was determined by the same method as above but the overlap of the images was performed on NDs on the substrate (cover slide) [10]. Then the displacement of the ND from the indentation position was read from the confocal image with optical resolution (∼300 nm).

*Synchronization:* In the AFM-confocal correlation microscope, we used NIDAQ (PCIe-6363, National

Instrument) for counting ND fluorescence and High Speed Data Capture (HSDC) (Nanoscope, Bruker) for AFM data acquisition, both with time resolution of 50 μs. The fluorescence and AFM data collections in the two microscopies were triggered by the same pulse signal sent out at the start of the force modulation (a negative-going pulse with width of 20 μs from the NanoScope V Controller of the AFM microscope, see Supplementary Figure S2 for the time sequences of the measurements). The synchronization of the two data collections were checked by collecting the fluorescence variation of an ND attached on the AFM cantilever under a non-contact height modulation of AFM tip (with the tip height acquired by AFM), as shown in Supplementary Figure S3. The lagging time between the two sets of data are less than 60 μs.

**B. Oscillatory rotation of magnetic field**

The coaxial oscillatory rotation of magnetic field was realized by applying a static magnetic field nearly in the *xy*-plane using a pair of permanent magnets, and a controllable magnetic field in the *z* direction by a magnetic coil with adjustable current (ranging between $\pm 0.5$ A), see Supplementary Figure S4(a) for details. The oscillation of the magnetic field in the *z* direction, realized by applying AC current to the magnetic coil, causes a rotation of the resultant external magnetic field applied on the ND, where the rotation axis is perpendicular to both the static magnetic field and the z-axis [see Supplementary Figure S4(b)].

Before each set of the indentation experiments, the external magnetic fields were calibrated using the ODMR spectra of the NV centers in a bulk diamond crystal. Two examples of the calibration are shown in Supplementary Figures S4(c) to (g). The rotation angle of the magnetic field, induced by the AC current in the magnetic coil, had the slopes of 19.4 deg./A and 13.4 deg./A in the PDMS and the live cell experiments [see Supplementary Figures S4(g)]. The variation of the magnitude of the external magnetic field during modulation ($\pm 0.2$ A) is less than 2% and is neglected [see Supplementary Figures S4(e)]. The polar angles of the projection of magnetic field on the *xy*-plane are 60.6 deg. and 90.2 deg. in the PDMS and live cell experiments, respectively.

**C. PDMS Sample preparation**

PDMS films were prepared using the Sylgard 184 elastomer kit (Dow Corning). The silicone base and crosslinker were thoroughly mixed with a weight ratio of 60:1 before degassing under vacuum. For AFM indentation experiments, the sample were prepared by spin-coating the mixture at 3000 RPM for 30 s onto a cover glass and cured at 60 °C for 24 hours. The thicknesses of the films are typically ~40 μm measured by confocal microscopy. The roughness of the PDMS surface is below 6 nm measured by AFM. 20 μL aqueous solution of ND with a concentration of 2 μg/mL (from Adámas Nanotechnologies, ND with ~900 NVs and diameter of 140 nm) were then drop-casted onto the PDMS film. The cover glass with the PDMS film was glued to a PCB board followed by soldering a microwave antenna (a copper wire with 20 μm diameter) on the surface of the sample connecting the microwave transmission lines on the PCB board. A confocal dish was then glued to the cover glass to form a liquid chamber with water added before experiments.

**D. Cell culture and incubation with NDs**

MCF-7 cells were seeded in a confocal dish and incubated at 37 °C with $CO_2$ level (5%) and humidity controlled. The dish was prepared in advance with a microwave antenna bonded to the surface of the cover glass. The growth medium of the MCF-7 cell was Dulbecco's modified Eagle's medium (DMEM, Gibco) supplemented with 10% fetal bovine serum (FBS, Gibco) and 1% penicillin–streptomycin (Gibco).

Cells were cultured for 48 hours before introducing NDs. The NDs (from Adámas Nanotechnologies, ND with ~900 NVs and diameter of 140 nm) were dispersed in sucrose solution (5μg $mL^{-1}$ concentration) and incubated with cell for 20 min at 37 °C in the incubator. With NDs attached on cell membrane after incubation, the sample was washed with PBS for three times and supplied with fresh DMEM-FBS medium. The dish was mounted onto the stage of AFM, with 5% $CO_2$ supplied during the indentation experiments. For imaging of cell by confocal microscopy, the plasma membrane of cells was labelled by the CellMask$^{TM}$ Green Plasma Membrane Stain (Thermo Fisher).

During the indentation experiments on cells, complete spectrum ODMR was collected before each indentation. The ODMR spectra were observed to change with time in contrast to those obtained on dead cells (see Supplementary Figures S19 and S20), suggesting that the cell sample remained alive during the experiment.

**E. AFM nanoindentation**

A DNP-10-A/PFQNM-LC-A tip was used in the experiments measuring the nonlocal deformation of PDMS/live cells upon AFM indentation (AFM: BioScope Resolve, Bruker). For a laterally homogeneous sample, its deformation resulting from an indentation at a fixed position can be effectively assessed by measuring the rotation of a single ND during multiple indentations conducted around the ND [10]. Consequently, we performed sequential AFM indentations near ND for nonlocal deformation mapping. For each indention, a loading force was applied and once the predetermined setpoint of the force is reached, a pulse signal was sent out from AFM triggering the start of the AFM modulation and ODMR measurements (the measurement duration is typically 25-30 seconds). The AFM modulation was implemented by the Ramp mode of the NanoScope software, where a modulation of tip height was added into the contact (hold) segment of the depth-loading curve.

We used AFM local depth-loading curves to check the uniformity of the mechanical response of the samples. The observed similar depth-loading curves at various locations on the sample surface around the ND verified the lateral homogeneity of the PDMS films (see Supplementary Figure S6) and MCF-7 cells (see Supplementary Figure S15).

**F. Analysis of the two-point ODMR and AFM data**

In each AFM indentation experiment, the ND fluorescence ($P_{\text{on}}(t)$ and $P_{\text{off}}(t)$ with the switching on- and off-resonance MWs of frequencies $\{f_i\}_{\text{on}}$ and $f_{\text{off}}$, see Supplementary Figure S7 for example) and AFM data (loading $L(t)$ and indentation depth $d(t)$) collections are triggered at the start of the force modulation, where a harmonic rotation of the external magnetic field are applied for calibrations (see Supplementary Figures S1, 2 and 4). The two-point ODMR signal is obtained as $S \equiv (P_{\text{on}} - P_{\text{off}})/P_{\text{off}}$ by normalizing the on-resonance fluorescence with the off resonance one [see Fig. 2(b) for example]. The effect of hydrodynamic drag on the AFM data was corrected by the drag constant in liquid. Moreover, the effect of cantilever reflection on the fluorescence signal was calibrated by applying an additional oscillation on the laser power. For details of the correction, see Supplementary Note 1 and Supplementary Figures S7 and S8.

In order to visualize the oscillation of the AFM and ODMR data by suppressing the shot noise, Figure 2(c) plots the result by separating the data (measured in 30 s) in continuous segments with duration of 0.2 s (with integer cycles of each oscillation) and then averaging all the segments. By applying the Fourier transformation, the AFM local data and the two-point ODMR signal were transformed to the frequency domain with frequency $f'$ as,

$$\tilde{x}(f') \equiv \frac{1}{T}\int_0^T x(t)e^{-i2\pi f' t}\, dt \tag{1}$$

where $\tilde{x}$ is a complex function and $T$ is the total measurement time as shown in Fig. 2(d) for examples. The amplitude and the phase of each modulation were extracted as the absolute value and the argument of the complex number $\tilde{x}(f)$. The amplitude of the ND harmonical rotation $|\tilde{\chi}(\rho)|$ [see Fig. 2(e) for example] was deduced by the ratio of the peaks of the loading to the magnetic modulations [see the bottom panel in Fig. 2(d)], and the known rotation angle amplitude of the magnetic field. The local (nonlocal) phase, i.e. $\phi_{\text{AFM}}$ ($\phi_{\text{ND}}$), was deduced by the difference of the phase between the loading and the depth (ND rotation) modulations.

**G. Viscoelastic model upon point loading with surface tension effect**

With the capillary effect included, an analytical solution was derived for the time-dependent nonlocal deformation $z(\rho, t)$, upon point loading in half-infinity incompressible linear viscoelastic model[27]. Under an oscillating point-loading with holding force as $L = L_0 + \delta L(t)$, the Fourier transformation of the deformation $z$ is written as,

$$\tilde{z}(\rho, f) = \frac{\tilde{L}(f)}{\pi} \int_0^\infty \frac{J_0(\lambda \rho)}{\tilde{E}^*(f) + 2\tau_0 \lambda} d\lambda \tag{2}$$

where $J_0(x)$ is the Bessel function, $\tilde{E}^*(f)$ is the complex modulus of the bulk with the phase $\delta_{\text{loss}}(f) = \arg[\tilde{E}^*(f)]$, and $\tau_0$ is the surface tension. The attached ND would be rotating following the deformation on the surface as $\tilde{\chi}(\rho, f) = \partial_\rho \tilde{z}$, where the modulation is written as

$$\tilde{\chi}(\rho, f) = \frac{\delta L(f)}{\pi \tilde{E}^*(f)} \int_0^\infty \frac{\lambda J_1(\lambda \rho)}{1 + \tilde{s}(f)\lambda} d\lambda \tag{3}$$

where $\tilde{s}(f) \equiv 2\tau_0/\tilde{E}^*(f)$ is the frequency-dependent elastocapillary length. After rescaling, the rotation modulation can be written as,

$$\frac{\tilde{\chi}(\bar{\rho}, \delta_{\text{loss}})}{\chi_L} = \frac{1}{\bar{\rho}} \int_0^\infty \frac{\lambda' J_1(\lambda')}{\bar{\rho} \exp(i\delta_{\text{loss}}) + \lambda'} d\lambda' \equiv \Theta(\bar{\rho}, \delta_{\text{loss}}) \tag{4}$$

where $\chi_L \equiv \delta L/(2\pi\tau_0|\tilde{s}|)$ is the normalization factor, $\bar{\rho} \equiv \rho/|\tilde{s}|$ is the rescaled distance. Supplementary Figure S9 plots the rescaled amplitude $|\Theta| = |\tilde{\chi}|/\chi_L$ and the normalized phase lag $\phi_{\mathrm{ND}}/\delta_{\mathrm{loss}} = -\arg(\Theta)/\delta_{\mathrm{loss}}$ as functions of the rescaled distance $\bar{\rho}$ at different $\delta_{\mathrm{loss}}$. The two functions are approximately universal functions with only weak dependences of the loss angle $\delta_{\mathrm{loss}}$. In the capillary limit where $\bar{\rho} \ll 1$, the normalized amplitude decays as $\rho^{-1}$ and there is no phase lag. In the viscoelasticity limit where $\bar{\rho} \gg 1$, the normalized amplitude decays as $\rho^{-2}$ and the phase lag is equal to that of the bulk. In the intermediate case, the decay behavior changes from $\rho^{-1}$ to $\rho^{-2}$, and the nonlocal phase lag increases from 0 to $\delta_{\mathrm{loss}}$ by increasing $\rho$ from the static capillarity-domaining region ($\rho \ll |\tilde{s}|$) to the dynamic viscoelasticity-domaining region ($\rho \gg |\tilde{s}|$) [see Fig. 2(g)].

**H. The evaluation of apparent complex modulus**

The indentation with a finite-size parabola tip into a homogenous viscoelastic material is simulated using the Hertz-Sneddon model [26]. In the oscillatory nanoindentation measurement, the indenter firstly indents to a settled depth $d_0$ and is derived by an oscillatory depth $\delta d$ with the frequency $f$. In the case that $\delta d \ll d_0$ [34], the apparent complex modulus at such frequency $\tilde{E}_a^*(f)$ can be evaluated by the FT data of the oscillatory depth $\tilde{d}(f)$ and loading $\tilde{L}(f)$ as,

$$\tilde{E}_a^*(f) = \frac{1}{2\sqrt{Rd_0}} \frac{\tilde{L}(f)}{\tilde{d}(f)}, \qquad (5)$$

where $R$ is the tip radius with $R = 25$ nm in Figs. 3(c) and 3(d) and $R = 65$ nm in Figs. 4(f) and 4(g).

**I. Oscillatory rheology measurement of PDMS**

The PDMS bulk sample was prepared by pouring the 60:1 mixture after degassing into a dish with 25-mm diameter to a thickness of ~3 mm and curing it at 60 °C for 24 hours (the same condition as stated above). The bulk mechanics of 60:1 PDMS was measured by the oscillatory rheology using a rheometer (MCR 301, Anton-Paar) equipped with a parallel-plate measuring system (PP25 with a diameter of 25 mm). The frequency sweeps were performed in the range of 0.1-20 Hz at a constant shear strain amplitude of 1%.

## ACKNOWLEDGEMENTS

This work was supported by RGC/GRF of HKSAR under projects no. C4007-19G, no. 14300720 and 14301722, Guangdong Provincial Quantum Science Strategic Initiative under Grant No. GDZX2205001, No. GDZX2305007 and No. QD2405001, the Science and Technology Development Fund (FDCT) of Macau under project no. 0122/2022/A, the Innovation Program for Quantum Science and Technology Project No. 2023ZD0300600, and Croucher Senior Fellowship 23401.

## AUTHER CONTRIBUTIONS

Q. L. and R. B. L. conceived the idea and supervised the project. W. H. L., Y. C., G. Z., R. B. L. and Q. L. designed the experiments. Y. C. and W. H. L. developed the spatial-temporal correlated measurements on AFM-ODMR system. W.H.L. designed the fast rotation tracking protocol and performed the numerical simulation. Y. C., G. Z., and W. H. L. performed the experiments. G. Z. and Y. C. prepared the samples. W. H. L., Y. C., G. Z., R. B. L. and Q. L. analyzed the data. Y. C., W. H. L., R. B. L. and Q. L. wrote the paper and all authors commented on the manuscript.

# Supplemental Material

*for*

# Nanodiamond-based spatial-temporal deformation sensing for cell mechanics

*Yue Cui et al.*

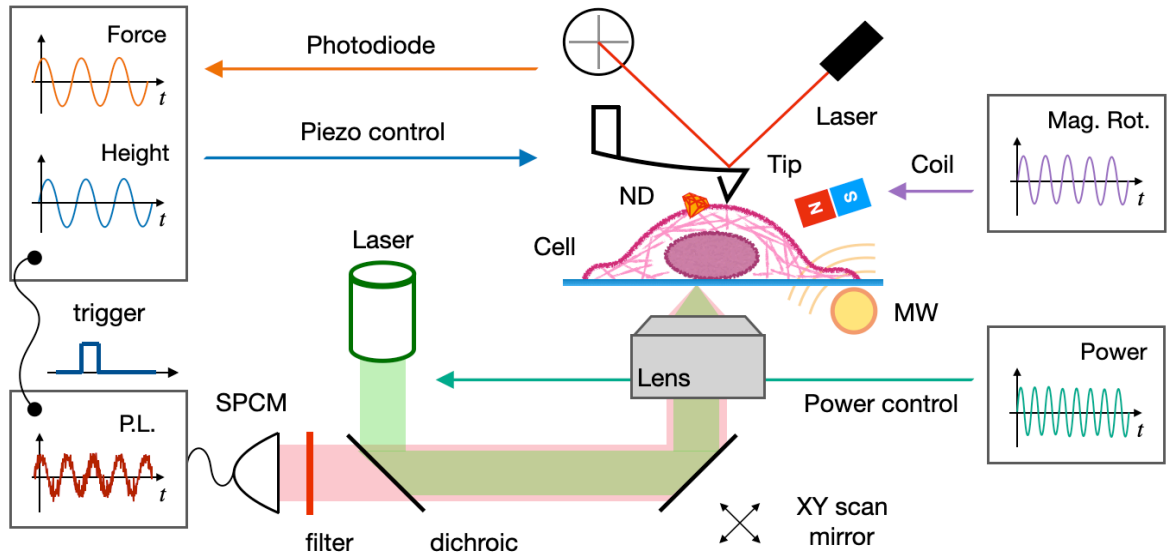

**Supplementary Figure S1 | Schematic of the confocal-atomic force microscope (AFM) setup.** *Confocal microscope:* a 532 nm laser was adopted (MGL-III-532-200 mW, CNI). A Nikon 100x (1.45 NA) oil immersion objective lens was used to collect the fluorescence of nanodiamonds (NDs), which was then detected by a single-photon counting module (SPCM, SPCM-AQRH-15-FC, Excelitas). *AFM*: the AFM scanning head (BioScope Resolve, Bruker) was mounted on the confocal microscope to measure the topography and to apply indentation. High Speed Data Capture (HSDC) function (Nanoscope, Bruker) was used for AFM data acquisition. *Optically detected magnetic resonance (ODMR) spectrum measurement:* Microwave (MW) sources (N5171B EXG Signal Generator, Keysight & WB-SG1-8G, Taobao) and an amplifier (ZHL-16W-43-S+, Mini-Circuit) were used to generate microwave (see Supplementary Figure S5 for details). A coil was put on the objective lens to rotate the electromagnetic field applied to the sample (see Supplementary Figure S4 for details). NIDAQ (PCIe-6363, National Instrument) counted the single-photon signal from the SPCM, and controlled the laser and the coil to modulate the laser power and the direction of the external magnetic field during indentation. Force modulation of the AFM tip was started by the NanoScope V Controller of the AFM microscope while a pulse signal (a negative-going pulse with width of 20 μs) was sent out and triggered the fluorescence and AFM data collections.

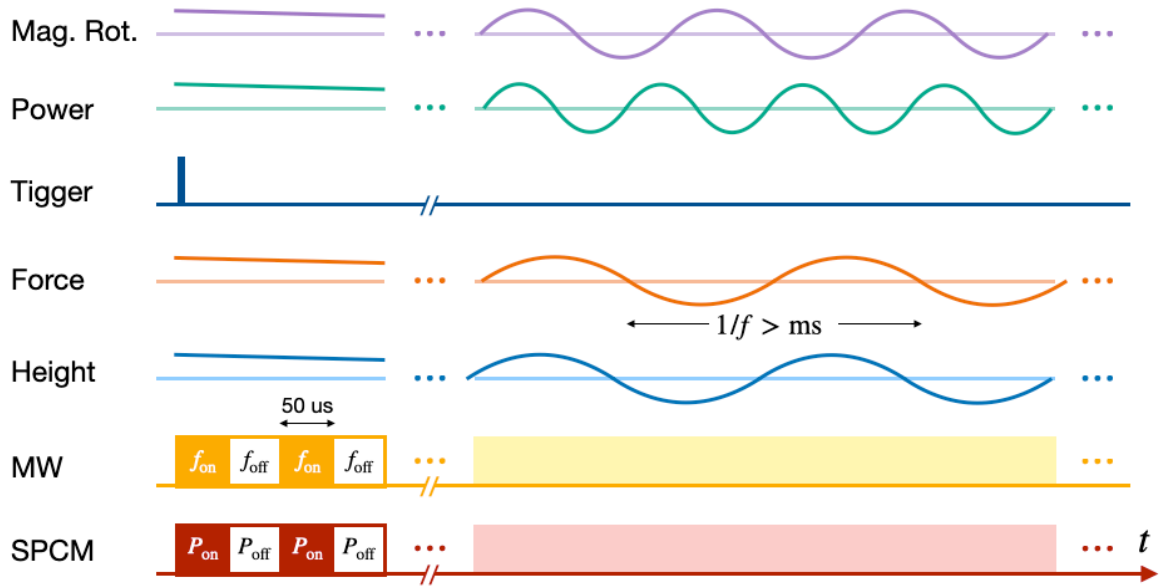

**Supplementary Figure S2 | Sequence for data collection in the dynamic nonlocal response measurement.** The modulation of the external magnetic field direction and laser power was consistently applied during the oscillating AFM indentation to calibrate the rotation angle magnitude of the NDs and correct for laser fluctuations (for more information, refer to Supplementary Note 1). The collection of AFM data, including force and corresponding tip heights measurements, was triggered by a pulse signal sent by the AFM, coinciding with the start of height modulation of the AFM tip. Simultaneously, the recording of ODMR signals was triggered by the same pulse signal, which alternated between on- and off-resonance microwave frequencies, $\{f_i\}_{on}$ and $f_{off}$ (see Supplementary Figure S5), and the corresponding fluorescence counting, $P_{on}$ and $P_{off}$, respectively.

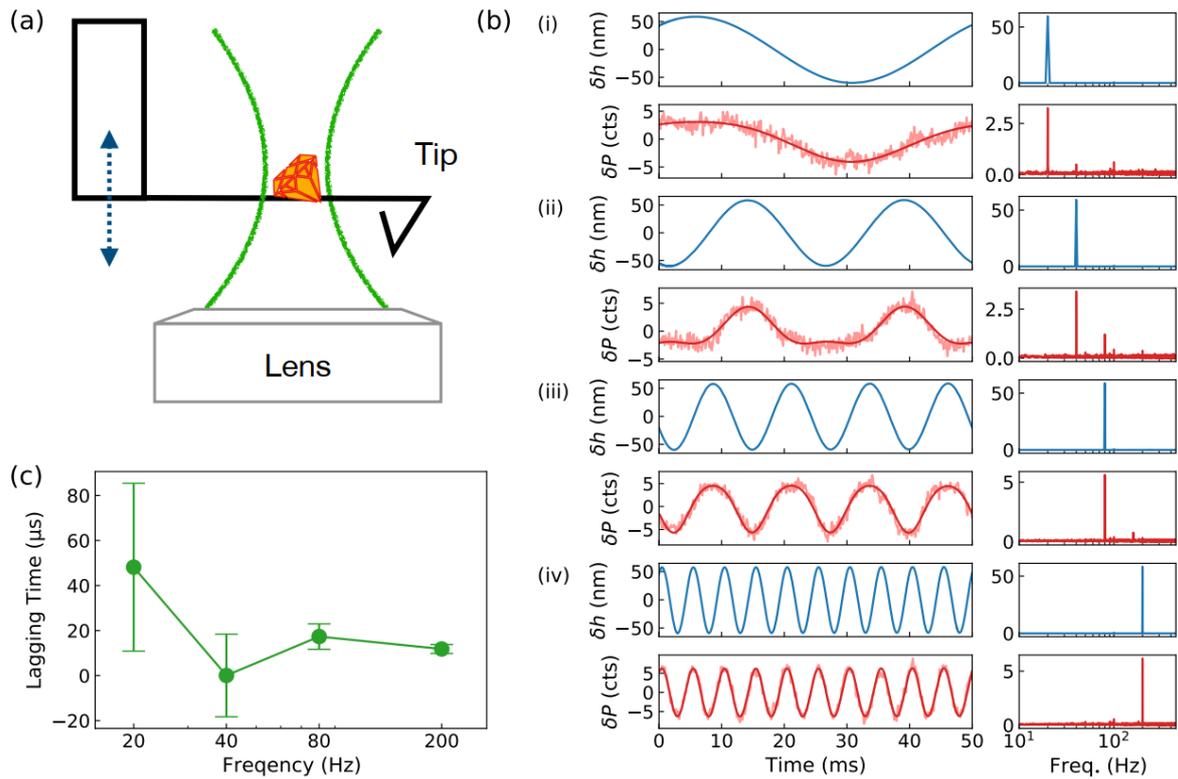

**Supplementary Figure S3 | Synchronization between the confocal microscope and the AFM.** (a) Scheme illustrating the preliminary experiment conducted to verify the synchronization between the confocal and AFM systems. NDs were drop-casted and positioned on the cantilever. Non-contact height modulations of the AFM tip were performed above the substrate, causing fluorescence oscillations of the ND that were captured by the confocal setup. (b) i to iv: The oscillating height data ($h$, captured by the AFM) and the ND fluorescence data ($P$, captured by the confocal setup) at different height modulation frequencies $f$ of 20, 40, 80, and 200 Hz, respectively. The Fourier transform results of the data are displayed in the right panel of (b). (c) The lagging time between the AFM (height data) and the confocal (fluorescence data) represented as a function of modulation frequency. The lagging time was calculated by multiplying the phase difference (obtained from the Fourier results) by the periodicity ($1/f$).

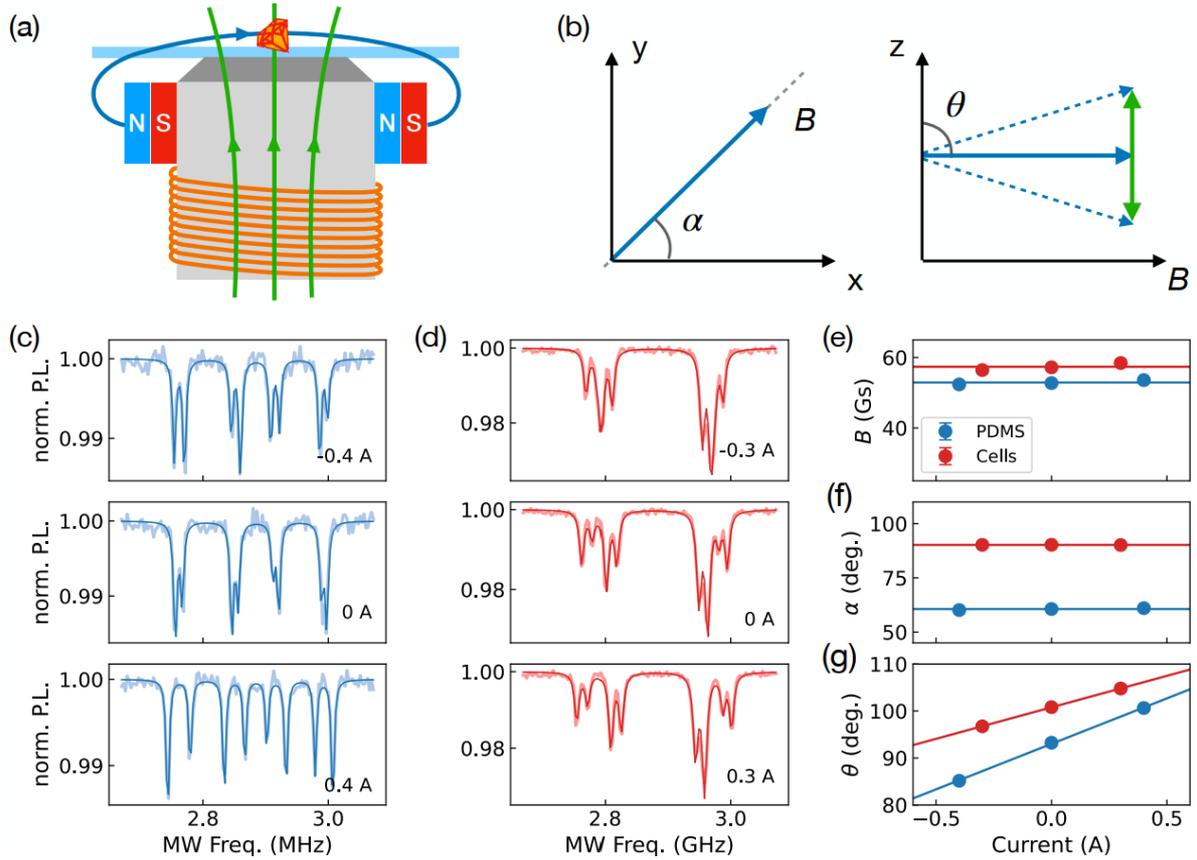

**Supplementary Figure S4 | Control and calibration of external magnetic field.** (a) The external magnetic field generated by a pair of permanent magnets and a magnetic coil fixed on the objective lens. The current through the coil is adjustable. (b) The static external magnetic field from the permanent magnets (blue arrows) and the adjustable field from the coil (the green arrows showing its range) add up to a total magnetic field that can be rotated in the range indicated by the blue dashed arrows. (c) and (d) The ODMR spectra obtained in the calibration of the external magnetic fields with a bulk diamond for the PDMS and the live cells experiments with different currents through the coil. The fitting of the ODMR spectrum under an external magnetic field is also plotted. (e) to (g) The magnitude $B$, azimuthal angle $\alpha$ and the polar angle $\theta$ as functions of the current though the coil. The magnitude and the azimuthal angle were approximately unchanged (with means of $\bar{B}$ =52.9 Gs and $\bar{\alpha} = 60.6 \ deg.$ for the PDMS cases; and $\bar{B}$ =57.4 Gs and $\bar{\alpha} = 90.2 \ deg.$ for the live cell cases), while the polar angle $\theta$ varied by the current with the slope of 19.4 $deg./A$ and 13.4 $deg./A$ in the PDMS and the live cell experiments, respectively.

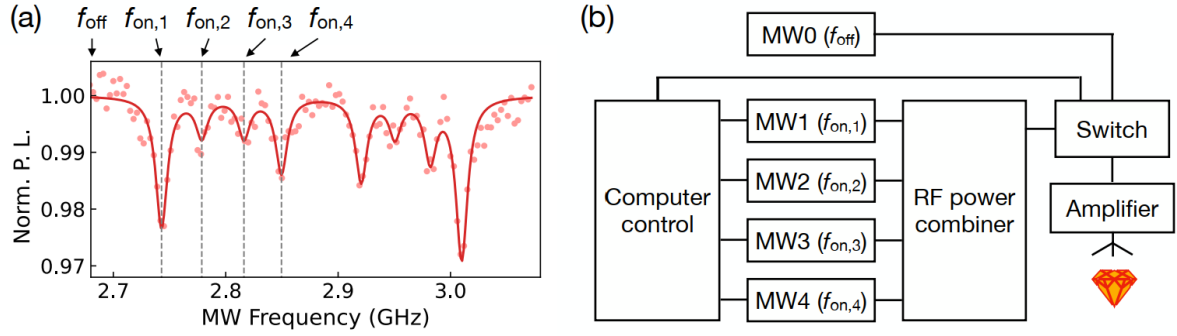

**Supplementary Figure S5 | Scheme of the four MW generators setup.** (a) A typical ODMR spectrum of the ND located on the PDMS as illustrated in Fig. 2(a) in the main text. The red line is the fitting results. The off-resonance frequency $f_{off}$ was set to 2.68 GHz. The dashed lines indicate the four on-resonance frequencies $f_{on,i}, i = 1\ to\ 4$, as estimated from the fitting results. (b) Scheme of the MW generators. The four MWs of on-resonance frequencies were mixed by a power combiner. A switch was used to switch the MW between the on- and off-resonance ones, and was controlled by the NIDAQ and the computer.

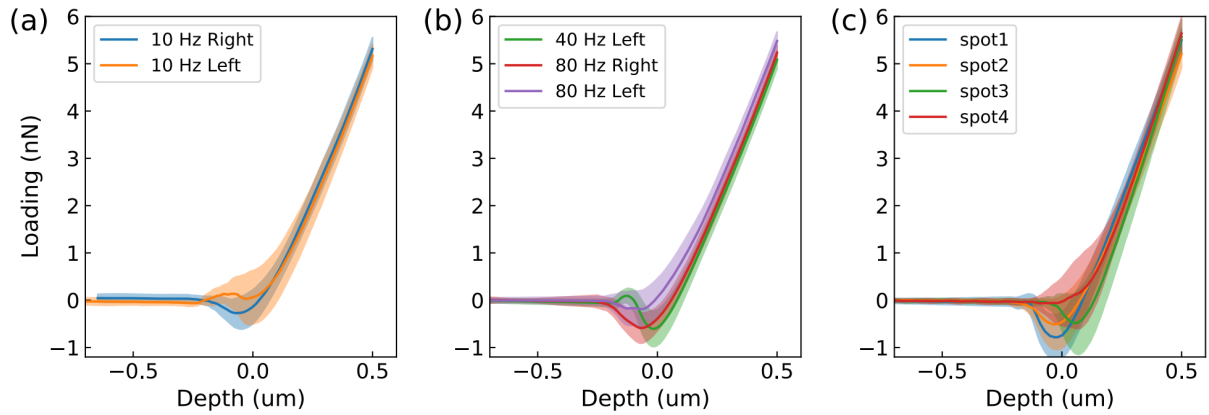

**Supplementary Figure S6 | Depth-loading curves of the AFM indentations on PDMS.** Averaged depth-loading curves of the local deformations under a constant indentation rate (600 $nm\ s^{-1}$) of (a) the 200 indentations as indicated in Fig. 2(a) in the main text, (b) the 250 indentations as shown in Supplementary Figures S10, and (c) the 320 indentations in Supplementary Figure S11. The error bars (given by the shadow) show the standard deviations of the measurements.

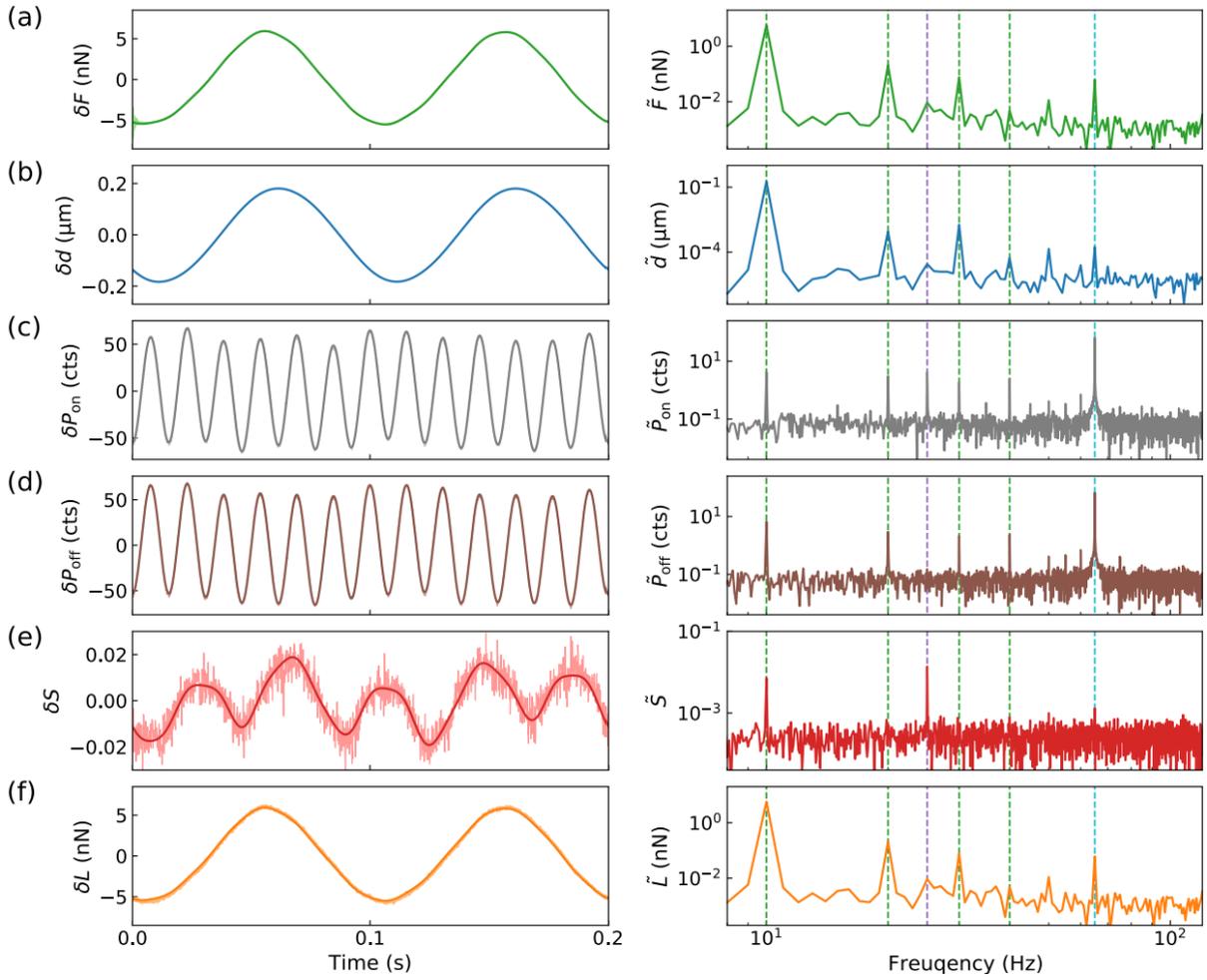

**Supplementary Figure S7 | Typical AFM and ODMR data of the dynamic nonlocal mechanical response measurement.** Left: The time averaging and Right: the Fourier transform (FT) of (a) the force $F$ and (b) the depth $d$ of the AFM tip, (c) the on resonant fluorescence $P_{on}$ and (d) the off resonant fluorescence $P_{off}$, (e) the two point ODMR signal $S$ and (f) the loading $L$ after the correction of hydrodynamic drag (for details, see Supplementary Note 1) at the indentation location indicated by the black arrow in Fig. 2(a) in the main text. The colored dashed lines in the right panel indicate the harmonical peaks of the FT signal induced by the oscillations of AFM tip (green), magnetic field (purple) and laser power (cyan).

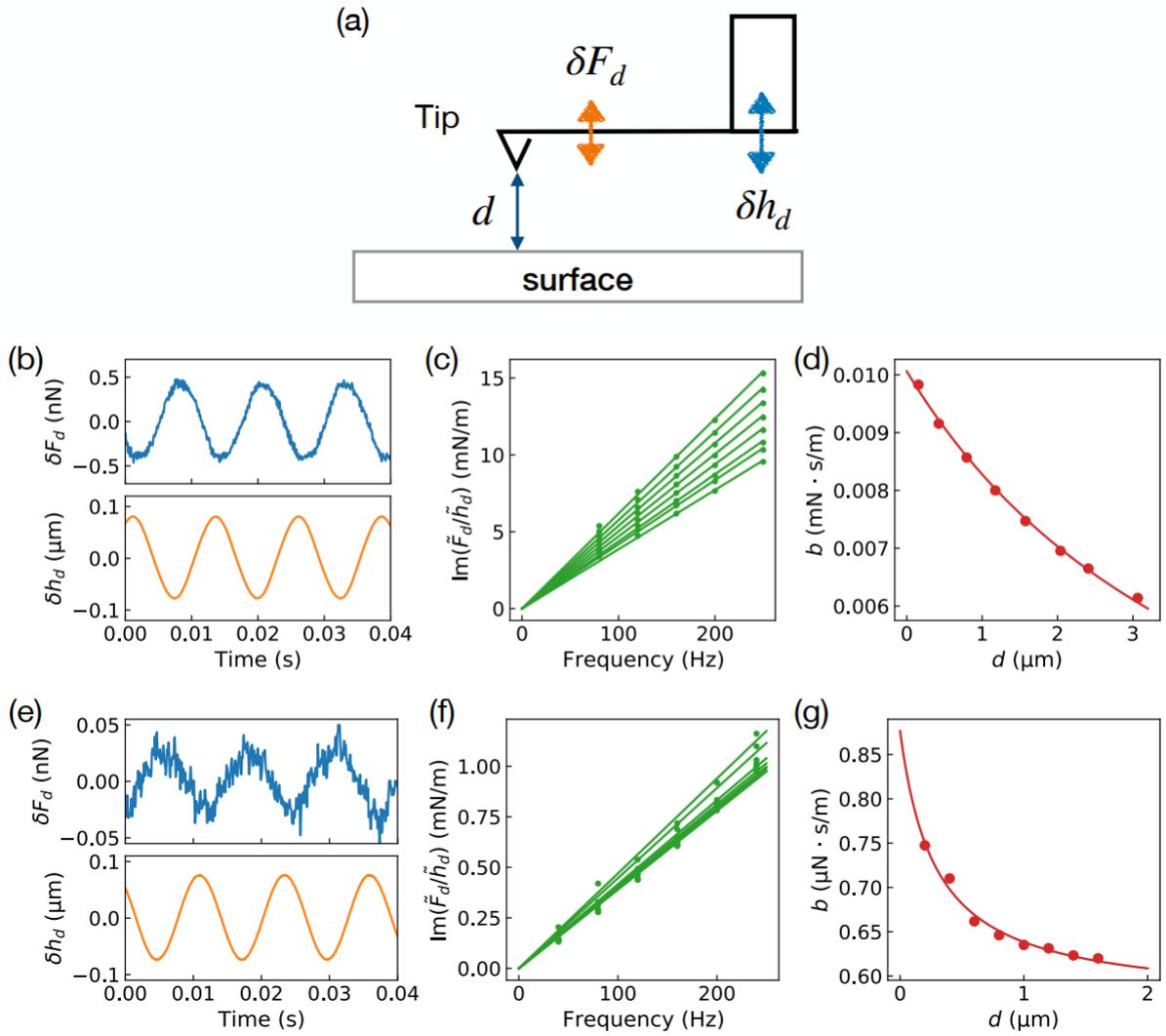

**Supplementary Figure S8 | Calibration of the drag constant of the AFM cantilevers.** (a) Scheme of the preliminary experiment for the drag coefficient calibration, see Supplementary Note 1 for details. (b) A typical drag force $F_d(t)$ and height modulation $h_d(t)$ with frequency of 80 Hz of the DNP-10-A tip (in the PDMS experiments). (c) The imaginary part of the ratio between the peak value of the Fourier transform of the drag force and the height as functions of the modulation frequency with different separation $d$ between the tip and the surface. The green lines are the linear fitting of the data. (d) the drag coefficient [slopes in (c)] as a function of the separation $d$. The red line is the fitting results. (e) to (g) the similar plots as those in (b) to (d) in the live cell experiments. The drag coefficients for the DNP-10-A tip (in the PDMS experiments) and PFQNM-LC-A-CAL tip (in the cell experiments) were calibrated to be $1.0 \times 10^{-5}$ and $8.7 \times 10^{-7}\ N\ s\ m^{-1}$ at $d = 0$, respectively.

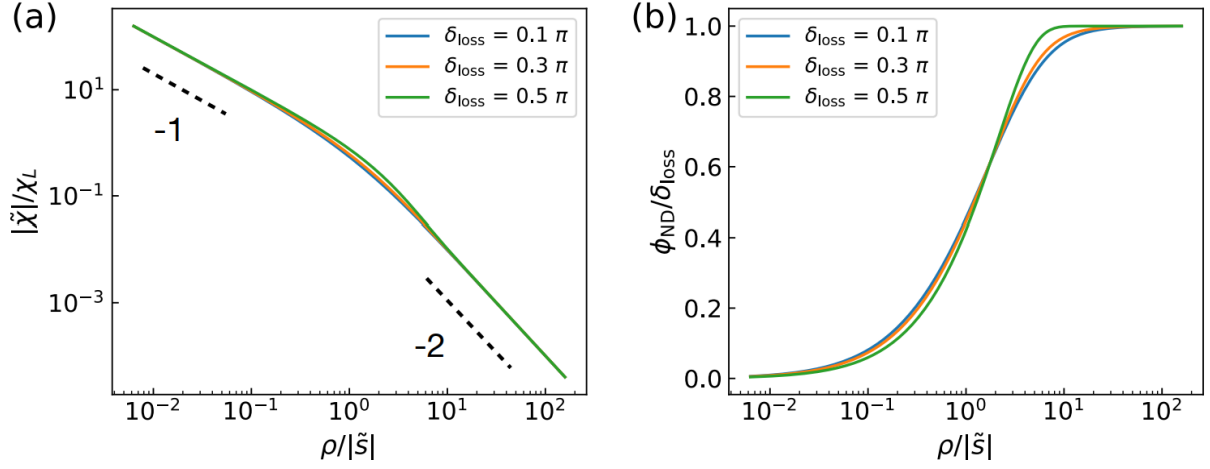

**Supplementary Figure S9 | The approximately universal functions of the viscous-elastocapillary model.** (a) The modulus $|\Theta| = |\tilde{\chi}|/\chi_L$ and (b) the normalized phase lag $-arg(\Theta)/\delta_{loss}$ of the approximately universal functions (see Methods in the main text) of the rescaled distance $\bar{\rho} = \rho/|\tilde{s}|$ at different $\delta_{loss}$. The dashed lines in a indicate the asymptotic behavior of $|\Theta| \approx \bar{\rho}^{-1}$ and $|\Theta| \approx \bar{\rho}^{-2}$ in the capillary ($\bar{\rho} \ll 1$, where the surface tension effect dominates) and the viscoelasticity ($\bar{\rho} \gg 1$, where the bulk effect dominates) limits.

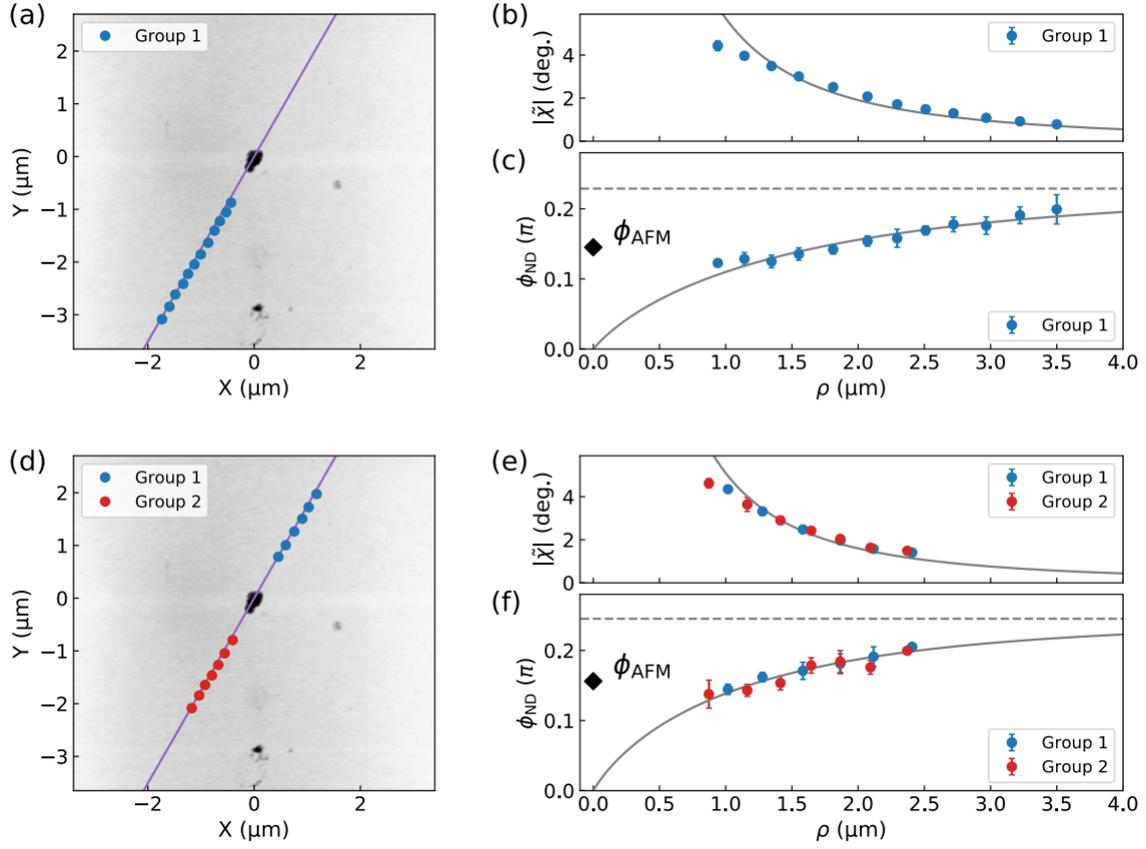

**Supplementary Figure S10 | The dynamic nonlocal response of PDMS under indentation at frequencies 40 & 80 Hz.** (a) AFM image as shown in Fig. 2(a) in the main text. The dots represent the indentation locations of the AFM tip around the ND. The purple line indicates the direction of the external magnetic field $\boldsymbol{B}$. (b) The amplitude of the oscillatory rotation angle ($|\tilde{\chi}|$) and (c) The nonlocal phase lag ($\phi_{ND}$) as functions of the distance $\rho$ between the indentation spot and the ND with modulation frequency of 40 Hz. The simulation results of the linear viscoelastic model including the surface tension effect are plotted by the grey lines. The local phase lag $\phi_{AFM}$ deduces from the AFM data is drawn by the rhombus at zero distance, while the bulk loss angle $\delta_{loss}$ is indicated by the grey dashed line. (d) to (f) the similar plots as those in (a) to (c) with modulation frequency of 80 Hz. The error bars in (b), (c), (e) and (f) are the standard derivations of the repeated measurements.

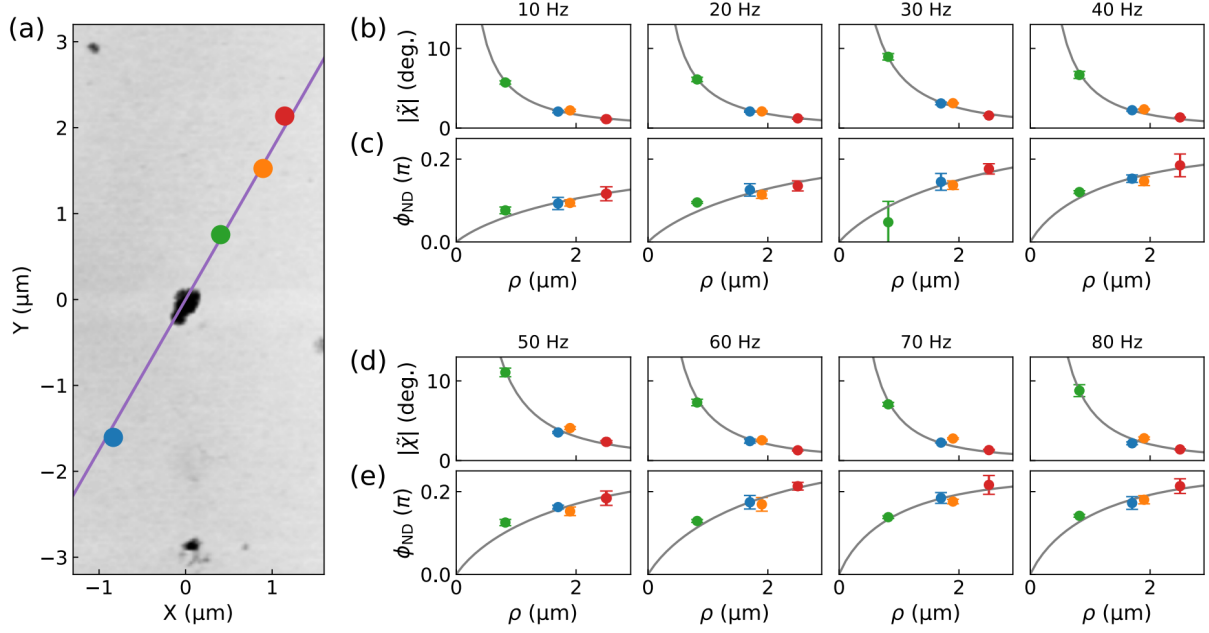

**Supplementary Figure S11 | The dynamic nonlocal response of PDMS with different indentation frequencies ranging from 10 to 80 Hz.** (a) The same AFM image of the PDMS surface as shown in Fig. 2(a) in the main text. The dots represent the indentation locations of the AFM tip around the ND. The purple line indicates the direction of the external magnetic field ***B***. (b) & (d) The amplitude of the oscillatory rotation angle ($|\tilde{\chi}|$), and (c) & (e), The nonlocal phase lag ($\phi_{ND}$) as functions of the distance $\rho$ between the indentation spot and the ND with different modulation frequencies ranging from 10 to 80 Hz. The simulation results of the linear viscoelastic model including the surface tension effect are plotted by the grey lines. The error bars in (b) to (e) are the standard derivation of the repeated measurements.

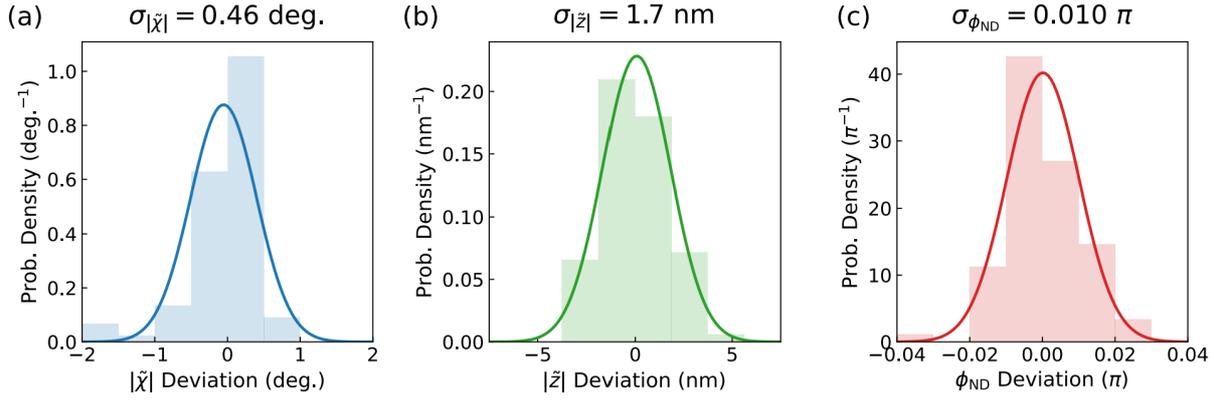

**Supplementary Figure S12 | Comparison between the experimental and simulation results of PDMS measurement.** The histogram of the deviations between the experimental and theoretical results of (a) the rotation oscillation amplitude $|\tilde{\chi}|$, (b) the non-local deformation amplitude $|\tilde{z}|$ and (c) the nonlocal phase lag $\phi_{ND}$ for the PDMS measured in Figs. 3(a) and 3(b) of the main text. The experimental results $\tilde{\chi}$ and $\tilde{z}$ were obtained by ODMR measurements and deformation reconstruction via integration (see Ref. [10] in the main text). The theoretical results (such as the grey lines in Figs. 3(a) and 3(b) in the main text) were calculated by using the viscoelastic model (see Methods in the main text). The solid lines are the fitting results with Gaussian distribution, with standard deviations shown at the top of each figure.

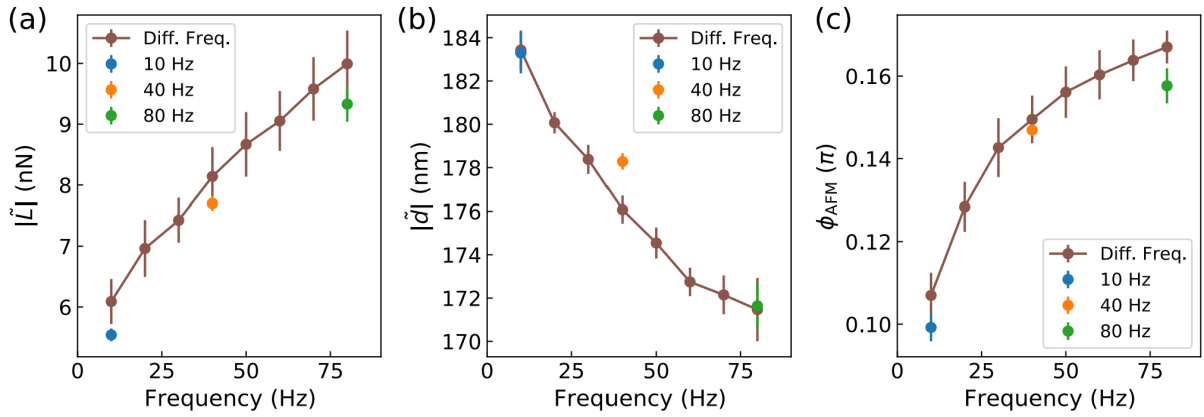

**Supplementary Figure S13 | The local indentation data of the PDMS.** The magnitude of the modulation of (a) the loading and (b) the indentation depth as functions of the modulation frequencies for the indentations shown in Fig. 2(a) in the main text (10 Hz), Supplementary Figure S10 (40 and 80 Hz), and Supplementary Figure S11 (10-80 Hz). (c) The corresponding phase lag $\phi_{AFM}$ between the loading and the depth of the indentations. The error bars are the standard derivations of the repeated measurements.

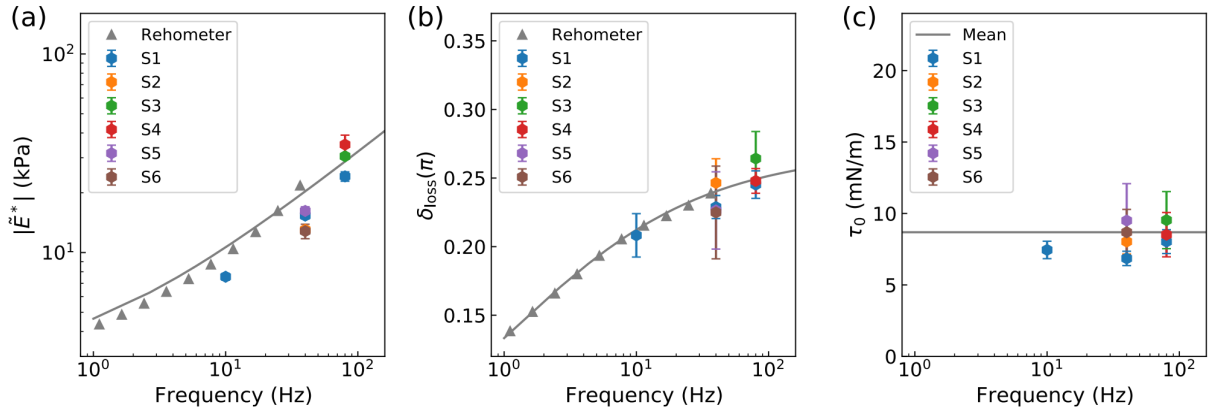

**Supplementary Figure S14 | The mechanical properties of different PDMS samples.** (a) The evaluated magnitude $|\tilde{E}^*|$ and (b) the loss angle $\delta_{loss}$ of the complex modulus as functions of the frequency $f$ for different PDMS samples. The blue dots (S1) are the complex modulus as shown in Fig. 3 in the main text. The grey triangles and lines plot the complex modulus of PDMS measured by rheometer and the corresponding fitting results evaluated by using the power-law function as shown in Fig. 3 in the main text. (c) The surface tension $\tau_0$ deduced in the different experiments. The line indicates the mean value. The error bars are fitting errors.

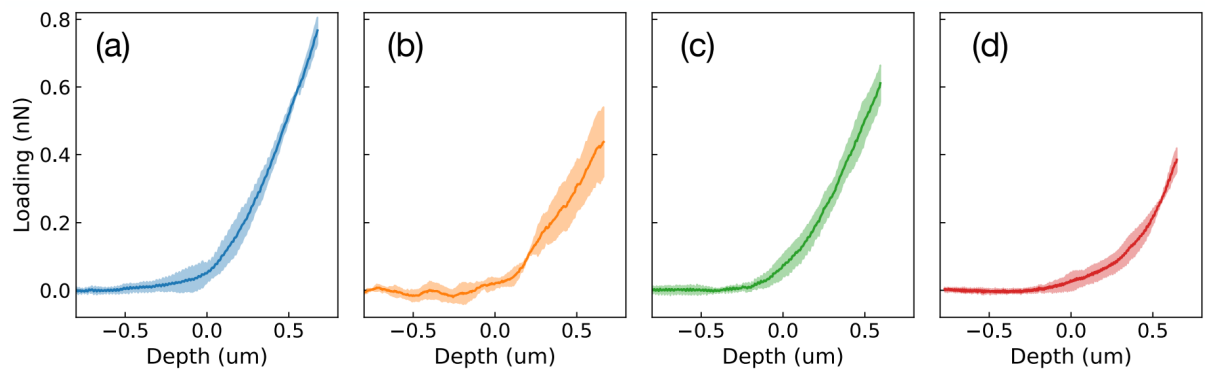

**Supplementary Figure S15 | Depth-loading curves of the indentation on the live cells.** (a) to (d) Averaged depth-loading curves of the local deformations under a constant indentation rate (600 $nm\ s^{-1}$) on the four live cells as illustrated in Fig. 4 in the main text. The error bars (given by the shadow) show the standard deviations of the measurements.

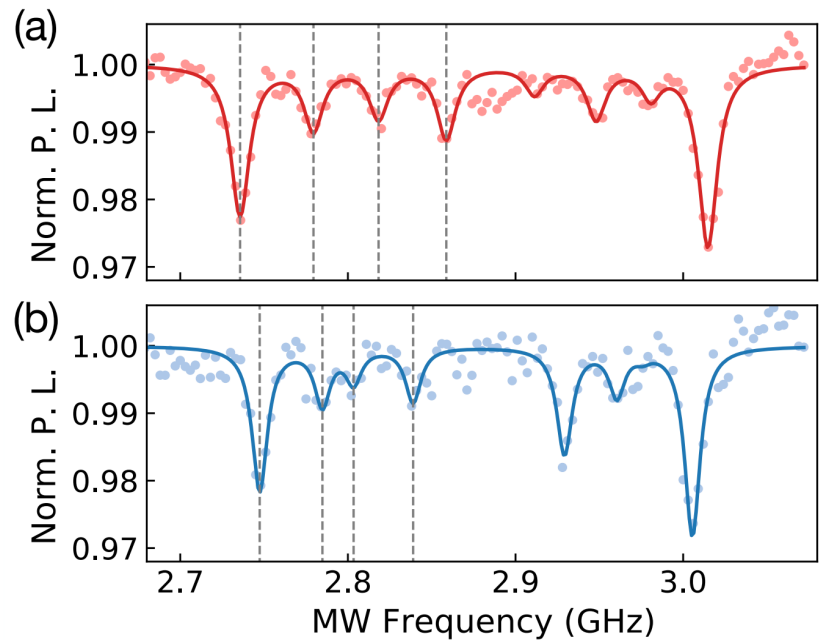

**Supplementary Figure S16 | Typical ODMR spectra obtained before each indentation.** (a) and (b) Two typical ODMR spectra of the attached ND obtained before the indentations on live cells. The colored lines are the fitting results. The grey dashed lines indicate the on-resonance frequencies of the transition between the $|0\rangle$ to $|-1\rangle$ electron spin states of the NV centers with the four directions. The four on-resonance frequencies used in the two-point ODMR measurement were determined by the ODMR spectra before each indentation.

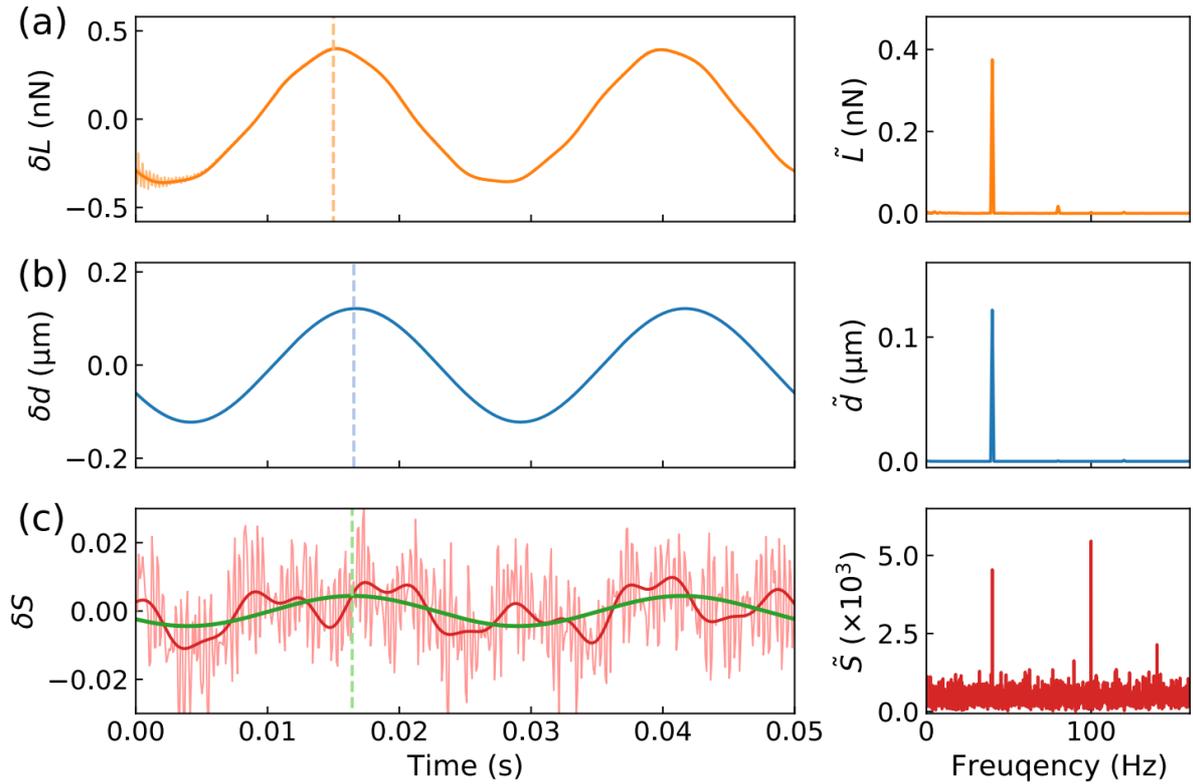

**Supplementary Figure S17 | The typical AFM and ODMR data of the nonlocal mechanical response measurement on live cells.** Left: The time averaging and Right: the Fourier transform (FT) of (a) the loading $L$ and (b) the depth $d$ of the AFM tip, and (c) the two point ODMR signal $S$ with a typical indentation on cell 1 in Fig. 4 in the main text. The dashed colored lines (orange, blue and green) indicate the phases of the respective oscillation data. The green curve in (c) is the signal from the ND rotation extracted from the two-point ODMR signal at the same frequency of the loading.

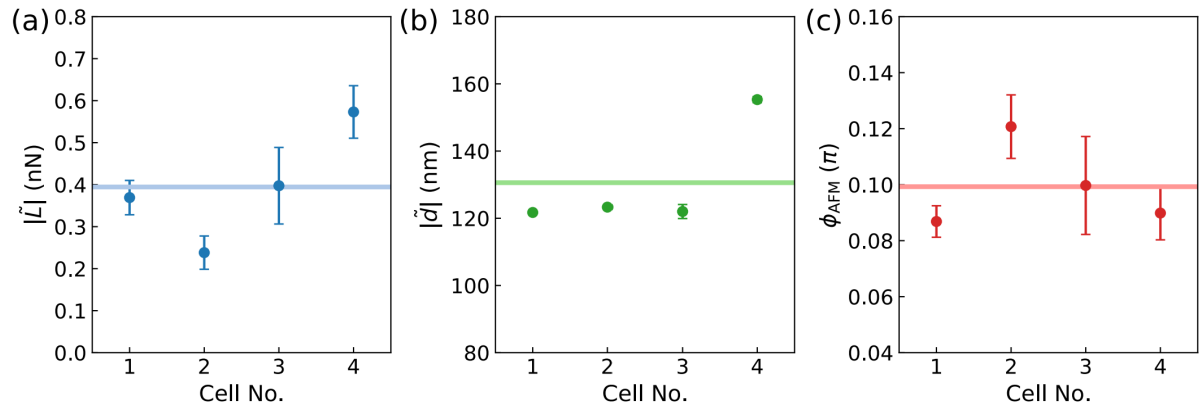

**Supplementary Figure S18 | The local indentation data on cells.** The amplitudes of the modulation of (a) the loading $L$ and (b) the indentation depth $d$ for the 40 Hz modulation on the four cells in Fig. 4 in the main text. (c) The corresponding phase lag $\phi_{AFM}$ between the loading and the depth of the indentations. The error bars are the standard derivation of the repeated measurements.

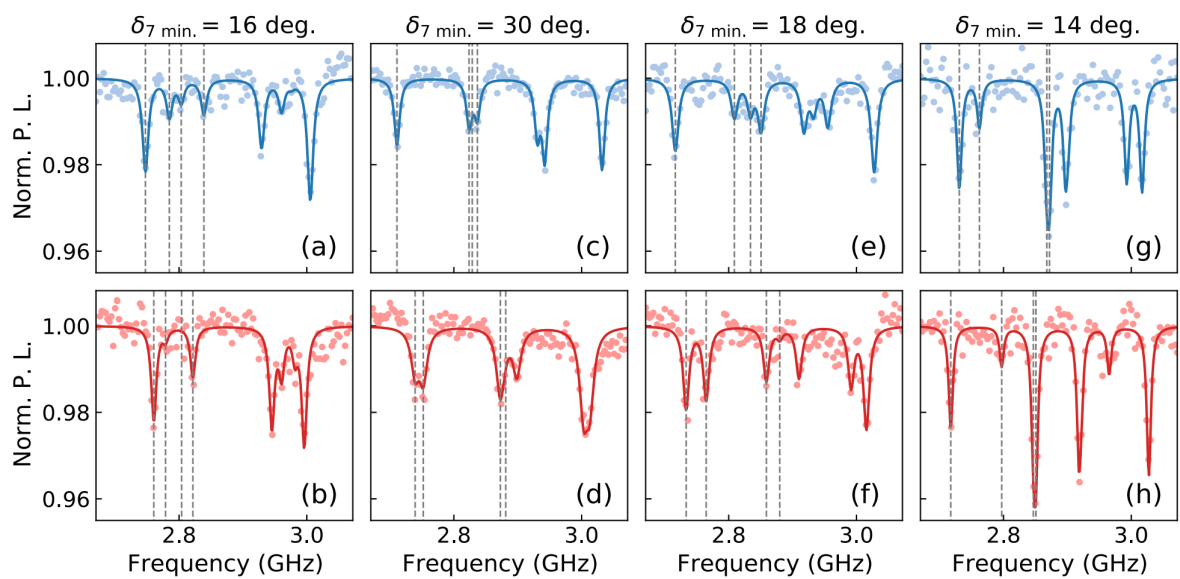

**Supplementary Figure S19 | The change in ODMR spectra of the ND attached on the live cells in a duration of 7 minutes.** (a) and (b) The ODMR spectra of the attached ND on the live cell 1 in Fig. 4 in the main text obtained after all the indentations. The time interval between the acquisitions of the two ODMR spectra was 7 minutes. The colored lines are the fitting results. The grey dashed lines indicate the on-resonance frequencies evaluated by fitting. The minimum rotation angle $\delta_{7\,min.}$ of the ND within the 7 min is estimated by the angle between the static external magnetic fields measured by the rotating ND [deduced by the fitting, similar to those in Supplementary Figures S4(c) and (d)] before and after the time interval. (c)&(d), (e)&(f) and (g)&(h) are similar to (a)&(b), but for the other three live cell experiments labelled as Cell 2-4.

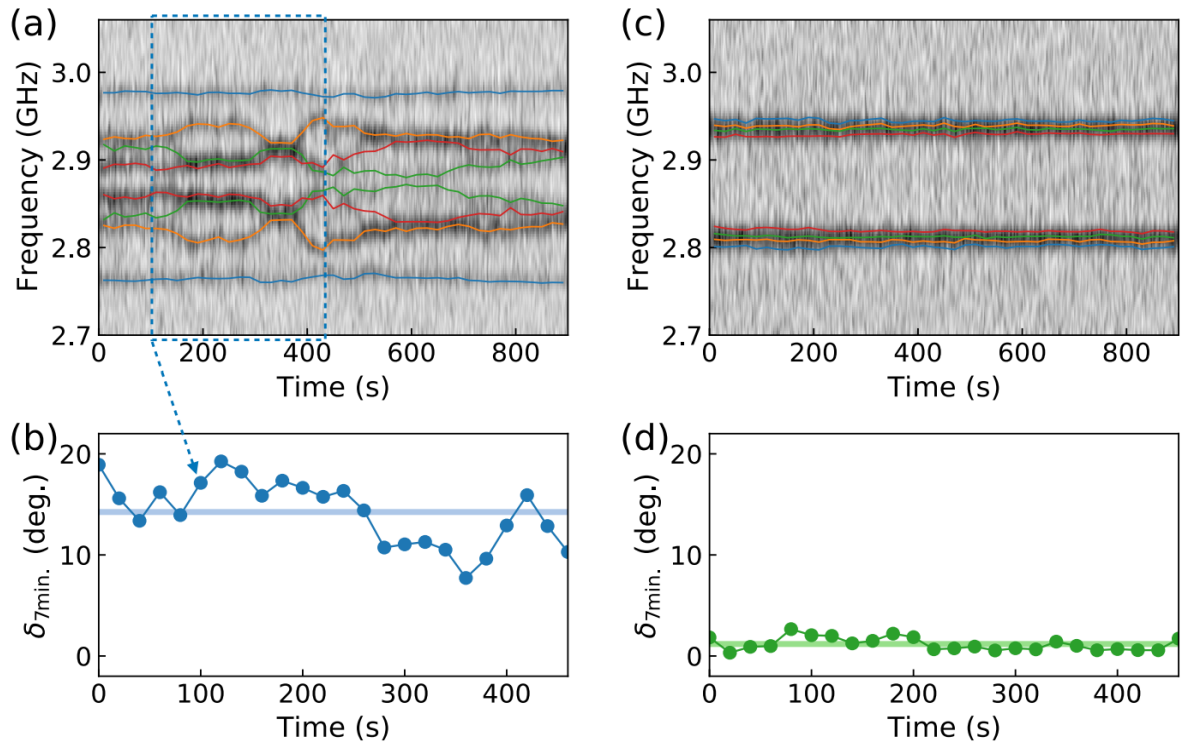

**Supplementary Figure S20 | The ODMR spectra of the ND on an MCF-7 cell being alive and dead.** (a) Time dependent ODMR spectra collected from the ND attached on the live MCF-7 cell under a static external magnetic field. The fitting results of the resonant frequencies are overlapped on the raw data as dashed colored lines. The contrast of the ODMR spectra is encoded in greyscale. (b) The minimum rotation angle $\delta_{7\,min.}$ of the ND within 7 min time duration (similar to those in Supplementary Figure S19, a typical time duration is labelled by blue dashed rectangular in (a) as a function of start time of each duration [indicated by blue dashed arrow in (a) for example]. The blue line is the mean value. (c) and (d) are similar to (a) and (b) but are the results measured after the cell died.

**Supplementary Note 1 | Analysis and correction of the two-point ODMR and AFM data**

In each AFM indentation experiment, the collection of the ND fluorescence data ($P_{on}(t)$ and $P_{off}(t)$ for MW frequencies being on-resonance $\{f_i\}_{on}$ and off-resonance $f_{off}$, respectively) and AFM data (force $F(t)$ and indentation depth $d(t)$) was triggered at the start of the force modulation, where an oscillatory rotation of the external magnetic field and an power modulation of the exciting laser were applied for calibrations (see Supplementary Figure S2 for the sequence of the measurements). Typical AFM and confocal data of an indentation in PDMS are shown as functions of time in the left panel of Supplementary Figure S7. The data (measured in 30 s) are displayed after averaging of the continuous segments with 0.2 s duration to suppress the shot noise of the fluorescence data. By applying the Fourier transformation, the AFM local data and the ND fluorescence (represented by $x(t)$) are transformed to the frequency domain with angular frequency $f'$ as,

$$\tilde{x}(f') \equiv \frac{1}{T}\int_0^T x(t)e^{-i2\pi f't}\,\mathrm{d}t \tag{S1}$$

where $\tilde{x}$ is a complex function and $T$ is the total measurement time. Examples of the data in frequency domain are shown in the right panel of Supplementary Figure S7. The colour dashed lines in the Supplementary Figure S7 indicate the first and high order modulation frequencies of force, magnetic field and laser power. Peaks are obtained in the AFM data at the force modulation frequency $f$ (10 Hz, indicated by green dashed lines) and the corresponding high-order harmonics ($2$ to $4f$, also the green ones), while multiple peaks are given in the fluorescence data corresponding to the modulations of the force ($f$), magnetic field ($f_B = 25$ Hz, the purple ones) and laser ($f_L = 65$ Hz, the cyan ones). Two corrections of the ODMR signals and the AFM signals were applied as follows.

*Normalization of ODMR signals:* By normalizing the on-resonance fluorescence with the off-resonance one, the two-point ODMR signals can be obtained as $S \equiv (P_{\text{on}} - P_{\text{off}})/P_{\text{off}}$. However, since the (linear) dependence of the fluorescence of NV centers on the laser power $p_L$ is different with the on- and off-resonance MWs applied (denoted as $\partial_{p_L} P|_{\text{on/off}}$) [35], the laser power fluctuation (including the one induced by the laser reflected to ND from the AFM cantilever with modulation frequency $f$) cannot be eliminated by the simple normalization. Hence, a modified normalization is applied as $S \equiv (P_{\text{on}} - \eta P_{\text{off}})/(\eta P_{\text{off}})$, where the factor $\eta$ is the ratio between the two linear dependences as $\eta = \partial_{p_L} P|_{\text{on}} / \partial_{p_L} P|_{\text{off}}$. To obtain this factor $\eta$, a laser modulation is applied in the ODMR measurement with frequency $f_L$ (see Supplementary Figures S1 and 2). The factor $\eta$ is then obtained as $\eta = |\tilde{P}_{\text{on}}(f_L)/\tilde{P}_{\text{off}}(f_L)|$, which is the ratio of the amplitude of the laser modulation peaks for the on-resonance to the off-resonance fluorescence (indicated by cyan dashed lines in Supplementary Figures S7(c) and (d). Supplementary Figure S7(e) shows the normalized two-po int ODMR signal. The correction on laser fluctuation is valid by the elimination of the peak at $f_L$ in the Fourier transform data ($\tilde{S}(f_L) \approx 0$, see the right panel).

*Hydrodynamic drag correction:* The force modulation captured by the AFM includes the loading $L(t)$ on the soft samples (i.e. PDMS and cells) and the hydrodynamic drag on the oscillating cantilever in liquid environments [7]. To eliminate the effect of hydrodynamic drag and extract the loading on the soft samples, we calibrated the drag coefficient in the PDMS and live cell experiments as illustrated in Supplementary Figure S8. In the calibration, the force response $\tilde{F}_d(f)$ of the cantilever to small height oscillations $\tilde{h}_d(f)$ at different frequency $f$ in the liquid environment approaching to (but not contacting) the sample with tip-sample separation $d$ was measured as schemed in Supplementary Figure S8(a). Typical AFM data are plotted in Supplementary Figure S8(b) for the DNP-10-A tip (in the PDMS experiments). The drag coefficient is deduced by the slope $\text{Im}[\tilde{F}_d(f)/\tilde{h}_d(f)]$ as a function of $f$ for different $d$ [see Supplementary Figure S8(c)]. Supplementary Figure S8(d) plot the drag coefficient as a function of the tip-sample separation $d$. Such a function was fitted by the scaled spherical model of the cantilever,

$$b(d) = \frac{6\pi\epsilon a_{\text{eff}}^2}{d + d_{\text{eff}}}, \tag{S2}$$

where $\epsilon$ is the dynamic viscosity of the liquid and $a_{\text{eff}}$ and $d_{\text{eff}}$ are the two fitting parameters accounting for the effective cantilever geometry. The red line in Supplementary Figure S8(c) is the fitting results. Similar calibration are shown in Supplementary Figures S8(e) to (g) for the PFQNM-LC-A-CAL tip (in the cell experiments). The drag coefficients $b_0$ for the DNP-10-A tip (in the PDMS experiments) and PFQNM-LC-A-CAL tip (in the cell experiments) were calibrated to be $1.0 \times 10^{-5}$ and $8.7 \times 10^{-7}$ N s m$^{-1}$ in the limit $d \to 0$. Hence, the loading of the indentation was obtained by the AFM data after correction as,

$$\tilde{L}(f) = \tilde{F}(f) - i2\pi f b_0 \tilde{d}(f). \tag{S3}$$

Supplementary Figure S7(f) shows a typical resultant loading $\tilde{L}(f)$ of the indentation.

The corrected loading $L(t)$, the indentation depth $d(t)$ and the normalized ODMR signal $S(t)$ are plotted in Fig. 2(c) in the main text after averaging. The amplitude and the phase of different modulations $x(t)$ were extracted as the absolute value and the argument of the complex number $\tilde{x}(f)$ (see Supplementary Figures S13 and S18 for typical loading and depth amplitudes in the PDMS and live cell experiments, respectively). The amplitude of the oscillatory ND rotation was deduced by

$$|\tilde{\chi}| = \left|\frac{\tilde{S}(f)}{\tilde{S}(f_B)}\right| |\tilde{\chi}_B|, \tag{S4}$$

where $|\tilde{\chi}_B|$ is the known amplitude of the magnetic field modulation controlled by the current of the coil (see Supplementary Figure S4). Finally, the local and nonlocal phases are evaluated as,

$$\phi_{\text{AFM}} = \arg\left[\frac{\tilde{L}(f)}{\tilde{d}(f)}\right] \quad \text{and} \quad \phi_{\text{ND}} = \arg\left[\frac{\tilde{L}(f)}{\tilde{S}(f)}\right], \tag{S5}$$